\documentclass[a4paper,fleqn]{cas-dc}

\usepackage[numbers]{natbib}
\usepackage{tabularx}
\newcolumntype{Y}{>{\centering\arraybackslash}X}

\usepackage{csquotes}
\usepackage{float}
\usepackage{listings}
  
\usepackage{tikz-uml}
\tikzumlset{fill usecase=white}
\tikzumlset{fill system=white}

\definecolor{dkgreen}{rgb}{0,0.6,0}
\definecolor{gray}{rgb}{0.5,0.5,0.5}
\definecolor{mauve}{rgb}{0.58,0,0.82}
\definecolor{aureolin}{rgb}{0.99, 0.93, 0.0}
\definecolor{bananayellow}{rgb}{1.0, 0.88, 0.21}
\definecolor{canaryyellow}{rgb}{1.0, 0.94, 0.0}
\definecolor{daffodil}{rgb}{1.0, 1.0, 0.19}
\definecolor{electricyellow}{rgb}{1.0, 1.0, 0.0}

\def\tsc#1{\csdef{#1}{\textsc{\lowercase{#1}}\xspace}}
\tsc{WGM}
\tsc{QE}
\tsc{EP}
\tsc{PMS}
\tsc{BEC}
\tsc{DE}

\begin{document}
\let\WriteBookmarks\relax
\def\floatpagepagefraction{1}
\def\textpagefraction{.001}
\shorttitle{Cataloging Dependency Injection Anti-Patterns in Software Systems}
\shortauthors{Laigner et~al.}

\title [mode = title]{Cataloging Dependency Injection Anti-Patterns in Software Systems}                      



\author[1]{Rodrigo Laigner}[type=editor,
                        orcid=0000-0003-2771-7477]
\cormark[1]
\ead{rnl@di.ku.dk}

\address[1]{University of Copenhagen, Denmark}
\tnotetext[1]{Work performed while enrolled as a graduate student in Pontifícia Universidade Católica do Rio de Janeiro, Brazil}




\author%
[2]
{Diogo Mendonça}
\ead{diogo.mendonca@cefet-rj.br}

\address[2]{CEFET/RJ, Brazil}

\author%
[3]
{Alessandro Garcia}
\ead{afgarcia@inf.puc-rio.br}
\address[3]{PUC-Rio, Brazil}



\author[3]{Marcos Kalinowski}[%
  ]
\ead{kalinowski@inf.puc-rio.br}

\begin{abstract}
\textit{Context:} Dependency Injection (DI) is a commonly applied mechanism to decouple classes from their dependencies in order to provide higher modularization. However, bad DI practices often lead to negative consequences, such as increasing coupling. Although white literature conjectures about the existence of DI anti-patterns, there is no evidence on their practical relevance, usefulness, and generality. \newline
\textit{Objective:} The objective of this study is to propose and evaluate a catalog of DI anti-patterns and associated refactorings. \newline
\textit{Methodology:} We reviewed existing reported DI anti-patterns in order to analyze their completeness. The limitations found in literature motivated proposing a novel catalog of 12 DI anti-patterns. We developed a tool to statically analyze the occurrence level of the candidate DI anti-patterns in both open-source and industry projects. Next, we survey practitioners to assess their perception on the relevance, usefulness, and their willingness on refactoring anti-pattern instances of the catalog. \newline
\textit{Results:} Our static code analyzer tool showed a relative recall of 92.19\% and high average precision. It revealed that at least 9 different DI anti-patterns appeared frequently in the analyzed projects. Besides, our survey confirmed the perceived relevance of the catalog and developers expressed their willingness to refactor instances of anti-patterns from source code. \newline
\textit{Conclusion:} The catalog contains DI anti-patterns that occur in practice and that are perceived as useful. Sharing it with practitioners may help them to avoid such anti-patterns, thus improving source-code quality.
\end{abstract}



\begin{highlights}
\item Misuse of dependency injection (DI) mechanisms leads to higher maintenance efforts
\item Literature presents no comprehensive characterization of bad DI implementation practices
\item We propose and evaluate a novel catalog of DI anti-patterns and refactorings
\item The proposed anti-patterns appear frequently in both open-source and industry projects
\item Expert practitioners confirm the relevance and usefulness of the catalog
\end{highlights}

\begin{keywords}
dependency injection \sep dependency inversion \sep inversion of control \sep anti-pattern
\end{keywords}

\maketitle


\section{Introduction}

Research in software engineering has been investigating several approaches for decreasing coupling in software systems over time, such as design patterns \cite{gamma:95} and aspect-oriented programming (AOP) \cite{kiczales:97}. Coupling is a quality attribute of a module in an application, and, high levels of coupling to other modules of the system tend to increase maintenance efforts \cite{gamma:95}. A particular mechanism that has been explored to decrease coupling levels in an application is dependency injection (DI)~\cite{prasanna:09}.

Dependency injection improves software modularity by enabling less coupling among modules by refraining them from being aware of implementation details (i.e., concrete implementation) of each other \cite{roubtsov:11}. Dependency injection is built upon two design principles: dependency inversion principle (DIP) \cite{martin:96} and inversion of control (IoC) \cite{fowler:04}. The first suggests a design oriented to abstractions (e.g., object-oriented interface), while IoC is about relying the control of the application to a third-party module or framework. Taken together, both design principles allow applications to evolve without incurring substantial dependencies across modules.

Given these benefits, dependency injection has become a common practice in the software industry, as characterized by the existence of a myriad of dependency injection frameworks, such as Guice \cite{guice:19} (Java) and Ninject (C\#), and industry-oriented publications \cite{prasanna:09,seemann:12}. For instance, Spring \cite{spring:19}, one of the most popular Java frameworks, Google AdWords\footnote{https://github.com/google/guice/wiki/AppsThatUseGuice}, a large-scale web application, and Microsoft Orleans\footnote{https://dotnet.github.io/orleans}, a framework for building scalable distributed applications, have their underlying components interconnected through dependency injection. Furthermore, due to the increased ubiquity of dependency injection usage in general-purpose software development, Java defined a specification targeted at dependency injection \cite{javax:01} and the recent Microsoft .NET Core already provides native dependency injection capabilities (i.e., no framework is necessary to activate dependency injection support as in earlier versions).


Despite the existence of well designed frameworks, such as Spring \cite{spring:19} and Guice \cite{guice:19}, that provide programming primitives to facilitate dependency injection usage, such as through code element annotations, the implementation of dependency injection is not trivial and demands in-depth knowledge on object-oriented design. Most importantly, improper dependency injection usage may actually hinder the effective achievement of its main goal, loose-coupling.

In this sense, although the technical~\cite{seemann:12,deursen:18} and white~\cite{roubtsov:11} literature conjecture about the existence of dependency injection anti-patterns and smells, these propositions do not provide a comprehensive analysis of the state of the practice of such bad dependency injection implementation characteristics in the source code of applications. For instance, the practical relevance of these propositions is unknown (e.g., the degree of occurrence in real projects and whether design principles are harmed), as well as their generalizability (i.e., whether those are generic enough to port its ideas to other contexts). Moreover, there is no evidence on the acceptance and perceived usefulness of existing proposals from the developers' point of view.


While properly employing dependency injection is an important step towards improving structural quality of software systems, there is a lack of guidance on how to effectively detect, analyze, and remove dependency injection anti-patterns from the source code. Consequently, practitioners have either little or no support on how to effectively handle DI anti-patterns in source code. Therefore, it is necessary to properly document dependency injection anti-patterns.

Based on this problem, using the design science template~\cite{wieringa:14}, our research goal can be defined as follows:

\textbf{Improve} the structural quality of software systems that employ DI 
\textbf{by} proposing and evaluating a catalog of DI anti-patterns
\textbf{that satisfies} providing comprehensive guidance on detecting, analyzing, and removing DI-related problems in software systems
\textbf{in order to} support practitioners in their development activities.

To meet our research goal, we answer the following research questions:
\begin{itemize}
  \item RQ1. Are there problem candidates associated with DI implementation that are not properly covered by the currently documented DI bad practices?
  \item RQ2. Do the proposed DI anti-patterns occur in practice?
  \item RQ3. Do expert developers understand the anti-patterns proposed as useful in practice?
  \item RQ4. Are practitioners willing to refactor DI anti-pattern instances from the source code?
\end{itemize}

To answer the research questions, we propose and evaluate a DI anti-pattern catalog using a quantitative study on the occurrence of the patterns followed by a qualitative study on acceptance and usefulness. This methodology, inspired by an explanatory sequential mixed-method design~\cite{ivankova2006using}, is detailed as follows.

\noindent\textbf{A. Proposing an initial catalog of DI anti-patterns.} First, we start by reviewing reported DI anti-patterns and other instances of problems related to DI employment in software systems. The objective is to analyze their completeness with the objective of uncovering gaps in current propositions in order to to come up with an initial catalog of DI anti-patterns. While industry-oriented publications mainly focus on .NET, we opted to target our investigation at the Java platform due to the following reasons: (a) the lack of documentation regarding bad DI implementation practices; (b) the existence of a myriad of DI frameworks (such as Guice \cite{guice:19} and Spring \cite{spring:19}); (c) large industrial adoption ; (d) the existence of a Java DI specification (JSR-330) \cite{javax:01}; (e) and the large number of open source software repositories written in Java.

We employ two methodological approaches to derive an initial proposition of DI anti-patterns, inductive and deductive. The inductive approach was primarily based on our experience in industrial settings. For the deductive approach, we conjectured on possible DI anti-patterns based on design principles, such as  General Responsibility Assignment Software Patterns (GRASP) \cite{larman:2004} and  SOLID \cite{martin2000design}. At the end of this process, RQ1 is answered.

It is noteworthy that, although our investigation comprises Java-based software systems, we believe that the catalog is generic enough to port its ideas to other languages in the same programming paradigm (i.e., object-oriented applications). Nevertheless, further investigation with this regard is important for such generalization and is out of the scope of this work.

\noindent\textbf{B. Investigating practical occurrence of the proposed catalog.} Next, after coming up with an initial catalog of DI anti-patterns, it is important to investigate the feasibility of our proposition. In other words, we seek to verify the practical relevance of the catalog by gathering the rate of occurrence of each anti-pattern instance in the context of both open and closed-source software projects. Therefore, we developed a static analysis tool to automatically detect instances of the proposed anti-patterns from the source code of software systems. We have selected a set of open-source software repositories from GitHub and two closed-source projects to perform our analysis on them. At the end of this process, RQ2 is answered.

\noindent\textbf{C. Investigating acceptance and usefulness of the proposed catalog.} Finally, we assess the acceptance and usefulness of our proposal from the point of view of expert developers by designing an expert survey. Although investigating the rate of occurrence of such proposed anti-patterns may lead to  practical relevance of the proposition, we still need to gather developers' perceptions over the proposed catalog. Specifically, we aim at gathering their perception over each proposed anti-pattern instance to verify if they can be characterized as anti-patterns. Besides, we assess whether the developers are willing to use our catalog to guide their development process. At the end of this process, RQ3 and RQ4 are answered.


As a result of applying this methodology, we list our contributions as following:

\noindent(i) We highlight the lack of comprehensive studies on dependency injection and analyze the limited relevance and generalizability of existing conjectured dependency injection anti-patterns, raising the need to properly document dependency injection anti-patterns;

\noindent(ii) We propose a novel catalog of dependency injection anti-patterns to support developers on avoiding the misuse of the dependency injection mechanism in software systems;

\noindent(iii) We develop a static analysis tool to detect instances of dependency injection anti-patterns in Java-based software projects that rely on DI for dependence resolution of its components. The tool demonstrates high precision and is available to the public on GitHub\footnote{https://github.com/rnlaigner/dianalyzer};

\noindent(iv) We show that the proposed dependency injection anti-patterns occur frequently within different open and closed-source software projects;

\noindent(v) We observed that practitioners perceive the proposed catalog as relevant and useful. Most importantly, they express willingness to refactor instances of dependency injection anti-patterns in source code, opening up venues for further investigation of the theme;

\noindent(vi) Through employing coding and categorization of respondents' opinions over each dependency injection anti-pattern of the catalog, we update and present a refined version of the catalog, further reflecting the industry needs.

It is noteworthy that this paper extends our previous conference version \cite{LaignerKCMG19}. Particularly, this article improves the previous version in seven aspects: (1) We fully document all twelve dependency injection anti-patterns, including source code samples, further details on the negative consequences, and the example solution for every dependency injection anti-pattern; (2) Regarding the investigation on the occurrence of each proposed dependency injection anti-pattern, we explicitly included the rules applied to the source code; (3) We include the results on the occurrence of dependency injection anti-pattern instances in two closed-source software projects obtained from industry representatives; (4) We included additional details about the expert survey execution; (5) We survey 15 additional practitioners (including 11 senior developers with experience in dependency injection implementation) to further assess the acceptance and usefulness of the proposed catalog; (6) In the new survey, we analyze an additional aspect about the willingness of the practitioners on refactoring instances of dependency injection anti-patterns, fomenting a new research question (RQ4); and lastly; (7) We employ a coding and categorization process to further characterize respondents' opinions over each dependency injection anti-pattern, thus fomenting a refined version of the catalog, which is more closely related to industry needs.

The remainder of this paper is organized as follows. Section 2 presents the background and related work. Section 3 describes how the DI anti-pattern catalog was proposed. Section 4 documents the proposed catalog of DI anti-patterns. Section 5 describes how we built and used a static analysis tool to assess the occurrence of the proposed DI anti-patterns in open and closed-source software projects. Section 6 presents our expert survey investigating the perceived usefulness. Section 7 summarizes how the catalog was updated based on the survey results. Section 8 discusses threats to validity. Finally, Section 9 concludes the paper.

\section{Background and Related Work}


This section provides a background on dependency injection and related work.

\subsection{Dependency Injection}
\label{subsec:dependency_injection}

According to Crasso et al. \cite{crasso:10}, dependency injection is a programming mechanism that "builds on the decoupling given by isolating components behind interfaces, and focuses on delegating the responsibility for component [or module] creation and binding to a dependency injection container". Besides, as noted by Yang et al.~\cite{yang:08}, dependency injection is a specific structural form of DIP. Indeed, dependency injection builds upon the dependency inversion principle, since components are decoupled through an interface oriented design. 

However, although one can achieve increased modularity through abstractions, it is still necessary to deal with the instantiation of concrete classes \cite{gamma:95}. The responsibility for component creation is then given to a dependency injection container (or DI container), a particular module employed in order to enable the IoC principle in dependency injection. Furthermore, a dependency injection container is the component of the dependency injection framework responsible for dependency provision at runtime, acting as a mediator in cases where a given dependence is required by a class.

For further details on dependency injection, including forms of dependency injection on runtime and the principles behind it, please consider Laigner's work~\cite{LaignerMaster20}.

Lastly, we believe it is worthy for the reader to consider referring to a terminology (Table~\ref{tab:terminology}) due to the extensive use of DI-related terms along this work.

\begin{table*}
  \caption{Dependency injection related terminology}
  \begin{tabularx}{\linewidth}{cX}
    \toprule
    \textbf{Term} & \textbf{Description}\\
    \midrule
    Injection & The act or process of dependence provision to a given application's component. \\
 \midrule
     Module & In this work, we refer to module interchangeably with a class in the object-oriented paradigm. \\
 \midrule
    DI Framework & A software framework employed to enable the dependency injection process and its benefits, offering code elements or structured file formats for the developer to specify in which points of the source code dependency injection is necessary. \\
 \midrule
     DI Container & An autonomous module (often provided by the DI framework) responsible for the injection of dependencies on application components. \\
 \midrule
    Injection Point & A developer-defined point of the source code entailed for receiving an injection of a dependence (e.g., usually through annotations in Java-based systems). \\
 \midrule
     Dependency Inversion Principle & The principle states that not modules should not depend on the details (e.g., concrete implementation) of another module, but rather abstractions (e.g., an interface) in order to allow for lower coupling and further reusing opportunities~\cite{martin:96}.\\
 \midrule
     The Principle of Inversion of Control & The principle is related to the ability of a software system to delegate parts of its own execution to an external agent or framework, which departs the system from controlling its entire execution life cycle~\cite{johnson:88}.
     \\
 \midrule
     Anti-Pattern & An anti-pattern is a set of characteristics found in elements of a software system (e.g., class, method, or attribute) that may lead to some sort of degradation, such as design and performance~\cite{brown:98}.\\
  \bottomrule
\end{tabularx}
\label{tab:terminology}
\end{table*}

\subsection{Related Work}
\label{subsec:related_work}

In this section, we discuss works that are related to the catalog of anti-patterns presented in this work.

\subsubsection{Dependency Injection}

Although there are some studies concerning the usage of dependency injection mechanism in applications, few of them concern cataloging anti-patterns of the dependency injection mechanism in applications as our work does.


\textbf{Dependency Injection Forms.} Yang et al. \cite{yang:08} conducted an empirical study concerning forms of dependency injection in Java applications and focused on analyzing projects that do not rely on a specific dependency injection framework. In order to measure the use of dependency injection, the authors defined four forms of employing dependency injection: constructor and method dependency injection, with and without default implementation. They employed a static analysis tool for finding these forms of dependency injection in 34 open source projects. The results show no evidence on the use of investigated forms of dependency injection and indicate that, instead, other mechanisms, such as service locators, were employed by the developers of the analyzed projects.

\textbf{Dependency Injection and Web Services.} Crasso et al. \cite{crasso:10} investigated the impact of DI on the development of web service applications in the context of DI4WS, a development model that allows for service discovery and consumption. The authors \cite{crasso:10} found out that DI enables faster development, cleaner code, looser coupling and simplifies service discovery, even though the overhead on memory allocation is higher compared to other design alternatives.

\textbf{Dependency Injection and Maintenance.} Razina and Janzen \cite{razina:07} conducted a study to measure the effects of the use of DI on software maintainability. In particular, the authors \cite{razina:07} focused on investigating the coupling and cohesion level of modules that apply the DI mechanism. They \cite{razina:07} selected a set of 20 open source systems, where each set contains a pair of software systems. The pair is composed by a project that employs DI mechanism and one similar project that does not employ DI. The authors \cite{razina:07} relied on three metrics to uncover the maintainability level of the projects: coupling between objects \cite{briand:99}, response for class \cite{briand:99}, and lack of cohesion in metrics \cite{chidamber:94}.

The authors \cite{razina:07} found that "[t]here does not appear to be a trend in lower coupling or higher cohesion measures with or without the presence of dependency injection." However, "a trend of lower coupling in projects with higher dependency injection percentage (more than 10\%) was evident."

Although the results exposed relevant findings, the validity of the work is questioned due to: (a) how the pairs of software systems were composed and (b) the characteristic (in terms of LOC and type of application) of the applications selected for the study.

\subsubsection{Catalogs of Code Smells}

Code smells are characterized as suspicious elements of code that may lead to a structural problem in the system~\cite{fowler2018refactoring}. Literature agrees that although a code smell does not decidedly yield a deeper structural problem, the odds on incurring in a problem are generally higher than the opposite \cite{cedrim:18}.

Although many studies have been carried out on code smells \cite{LACERDA2020110610,refactoring}, none of them have studied the specific context of dependency injection. Furthermore, the literature on code smells is oblivious to the dependency injection technique. Nevertheless, some DI anti-patterns presented in our catalog are related with code smells~\cite{fowler2018refactoring}. When necessary, we indicate such relation along the presentation of each anti-pattern in Section~\ref{sec:catalog}.

\subsubsection{Catalogs of Anti-Patterns}

Software engineering literature has explored catalogs with the objective to clearly communicate to practitioners best practices when it comes to the codification of software. Soon, software engineering researchers and practitioners realized that although there were clear guidelines on promoting good object-oriented design, applications still suffered from several problems, such as poor design, technical debts, and bugs~\cite{brown:98,larman:2004}.

In line with the words of Arnaoudova et al.~\cite{arnaoudova:13}, researchers became aware that, more than patterns, practitioners should be warned about what not to do or what practices to avoid when it comes to designing software projects. Thus, catalogs of anti-patterns have been designed in order to overcome these aforementioned problems. A set of works that introduce anti-patterns in software engineering (in areas related to linguistics~\cite{arnaoudova:13}, logging~\cite{chen:17}, and infrastructure as code~\cite{RahmanFW20}) and were influential to this work. However, none of them targets dependency injection concerns.

\subsection{Wrapping it up} 

The studies presented in this section analyze aspects related to the use of forms of enabling dependency injection at runtime~\cite{yang:08} and the effects of DI on development~\cite{crasso:10} and maintainability~\cite{razina:07} of software systems. However, we observe that existing empirical and mining studies on structural quality lack extensive discussion over DI bad practices.

In the next section, we turn our attention into the technical literature that suggests the existence of bad practices related to the use of dependency injection, such as smells and anti-patterns. We discuss their incompleteness and pave the way for a novel catalog of dependency injection anti-patterns.

\section{Proposing a Catalog of Dependency Injection Anti-patterns}
\label{sec:proposal}

Software engineering researchers have carried out many studies on improving structural quality of software systems \cite{cedrim:18}. However, we observe that existing empirical and mining studies on structural quality lack in-depth discussion over DI anti-patterns. The studies analyze aspects related to the use of forms of DI \cite{yang:08}, and the effects of DI on development \cite{crasso:10} and maintainability \cite{razina:07} of software systems.

We start this section by reviewing existing propositions of bad practices related to dependency injection usage in software projects.

\subsection{Analyzing Existing Propositions}

\label{sec:di_bad_smells}
\textbf{Dependency Injection Bad Smells}. Roubtsov et al. \cite{roubtsov:11} claim that overuse of annotations can potentially lead to violations of modularity principles. They propose a catalog of "bad smells" over dependencies injected in the context of Java annotations. A summary of the proposition is shown in Table \ref{tab:di_bad_smells}. The first column represents the modularity principle behind the bad smell and the second column exhibits the annotations involved in the respective violation. 
\begin{table}
  \caption{Summary of DI bad smells from Roubtsov et al. \cite{roubtsov:11}}
  \begin{tabularx}{\linewidth}{XX}
    \toprule
    \textbf{Principle} & \textbf{Annotations}\\
    \midrule
    Configuration should be separated from functionality  & \makecell[l]{Application server: \\ \textit{@Install} and \textit{@Startup} \\
Web server: \\ \textit{@Path} and \textit{@RequestMapping} \\
Database Server: \\ \textit{@Table} and \textit{@Column}}
 \\
 \midrule
Information should not be duplicated & Principle violated by interface annotations mentioning the interface implementation 
(@ImplementedBy, @ProvidedBy) \\
   \midrule
  Information should not be duplicated & Annotations duplicating the database structure (@Id, @OneToOne, @OneToMany, @ManyToOne, and @ManyToMany) \\
  \midrule
  Interfaces should not be explicit & Java interceptor annotations (@Interceptors, @AroundInvoke) \\
  \bottomrule
\end{tabularx}
\label{tab:di_bad_smells}
\end{table}

The authors \cite{roubtsov:11} assert that annotation \textit{@ImplementedBy} triggers a potential inconsistency due to maintenance. It is important to address that this impact is only achieved in case of the introduction of another implementation possibility. In addition, Roubtsov et al. \cite{roubtsov:11} argue that circular dependency can be achieved between interface and its implementation. Although the authors \cite{roubtsov:11} did not provide an example, design patterns, such as Factory, can be used as mechanisms for solving circular dependencies.

Regarding configuration annotations, such as \textit{@Install} and \textit{@Startup}, [26] address that separating application server configuration and build files is a feasible resolution. It is important to observe that configuration annotations are mechanisms by which frameworks can introduce its own behavior in the application on run time. Then, coupling in this context cannot be excluded. However, it is possible to provide an interface oriented design, isolating classes annotated with these configuration annotations, on which the binding of the instance is accomplished by a DI container. Isolating these classes in a different component is in charge of developer, being a pure architectural choice.

Over dependence on the web-server violation, on which the authors \cite{roubtsov:11} claim that "redeployment of the software on a new server is hindered by presence of explicit dependencies on web-pages", web-deployment descriptors (configuration files) is not the only solution. It is noteworthy that annotations mainly goal is to diminish the need for configuration files. 

Lastly, about annotations concerning database structure, we assert that due to Java Persistence API, a programming interface specification, Roubtsov et al. \cite{roubtsov:11} argument over the impact of redeploying the Java system on a new server due to presence of explicit dependencies on the database tables, is not conceivable. As a pattern on the Java platform (Java persistence API), redeployment in a new server is unlikely to require the removal of persistence API annotations.

\label{sec:di_anti_patterns}
\textbf{Dependency Injection Anti-Patterns}. The only explicit proposal of DI anti-patterns is the one described by Seemann \cite{seemann:12} and Deursen and Seemann \cite{deursen:18}, which contains a set of four DI anti-patterns, shown in Table \ref{tab:current_di_ap}.

\begin{table}
  \caption{DI anti-patterns extracted from Seemann \cite{seemann:12}}
  \begin{tabularx}{\linewidth}{cX}
    \toprule
    \textbf{Name} & \textbf{Description}\\
    \midrule
    Control Freak & Dependencies are controlled directly, as opposed to Inversion of Control \\
 \midrule
    Bastard Injection & Foreign defaults are used as default values for dependencies \\
 \midrule
    Constrained Construction & Constructors are assumed to have a particular signature \\
 \midrule
    Service Locator & An implicit service can serve dependencies to consumers but isn't guaranteed to do so \\
  \bottomrule
\end{tabularx}
\label{tab:current_di_ap}
\end{table}



In \textit{Control Freak} anti-pattern, the principle of inversion of control is not achieved, since a dependence is obtained through directly creating an instance of a concrete implementation. This behavior introduces high coupling in the system through modules that make use of direct creation of instances. 

\textit{Bastard Injection} concerns specifying a default constructor with the objective of creating a default dependence instance. Usually implemented aiming at supporting unit testing, the class with a bastard injection incurs in high coupling with the default dependence created by its default constructor.

\textit{Constrained Construction} regards introducing an implicit constraint on a dependence, i.e., a constructor with a particular signature. This behavior represents a problem when late binding is necessary due to application requirements.

Finally, \textit{Service Locator}, a design pattern introduced by Martin Fowler \cite{fowler:04}, is described by Seemann \cite{seemann:12} as an anti-pattern in the context of DI applications, since it implies in widespread coupling to a static factory (in this case, the Service Locator class) throughout source code.

\textbf{Wrapping it up.} Regarding the work of  Roubtsov et al. \cite{roubtsov:11}, even though the authors provide means for resolving each smell, they provide no comprehensive discussion concerning the validity of the proposed "code smells". It is important to note that, from thirteen annotations analyzed, only two are related to DI (\textit{@ImplementedBy}, \textit{@ProvidedBy}), which are annotations introduced by the Guice framework. Moreover, they recognize that the cataloged smells heavily focused on annotations related to the J2EE persistence model, which are based upon Java Persistence API (JPA), a specification for persistence in Java.

Furthermore, the DI anti-patterns addressed by Seemann \cite{seemann:12} and Deursen and Seeman \cite{deursen:18} correspond to generic rules of thumb when it comes to DI adoption in software projects. For instance, regarding \textit{Control Freak}, the author suggests that the presence of the \textit{new} keyword and static factories as indicative of high level of coupling in source code. However, \textit{Control Freak} can also be considered an anti-pattern in projects that do not implement or follow DI principles. Furthermore, the same can be said about \textit{Bastard Injection} and \textit{Constrained Construction}. The first may also harm modularity even though no DI framework is used in the application. The second even though the constructor enforce a set of dependencies, in the absence of concrete details (i.e., the constructor parameters are interfaces), it does not necessarily harm DI principles.

The only proposition that decidedly violate inversion of control is the \textit{Service Locator}, which takes the role of directly providing concrete instances over different modules in the system.



Therefore, given the limitation of current propositions of bad practices in dependency injection to address design principles behind DI, it is important to effectively investigate and characterize elements of source code that decidedly hinder the proper employment of DI in software projects. This work, through a proposal of a novel catalog of DI anti-patterns and subsequent investigation of its practical relevance, intends to fill this gap. The next section present the efforts to fulfill these identified gaps.

\subsection{Proposing a Novel Catalog of Dependency Injection Anti-patterns}

As mentioned above, previously reported DI anti-patterns aim at generic problems. For instance, \textit{Control Freak} can also be found in other contexts where inversion of control is adopted without a DI framework. In addition, we were not able to identify existing literature regarding DI anti-patterns in the context of Java. Lastly, reported DI anti-patterns and DI code smells fail to depict their application context scenario, and most importantly, fail to present evidence on their practical relevance.

In order to address such limitations, in this section, we report on our efforts in documenting a candidate catalog of DI anti-patterns. Although we believe the DI anti-patterns proposed are generic enough to port its ideas to other object-oriented programming languages, such as C\# and Python, in this proposal, we opted for drawing on the Java platform due to the following reasons: (a) the existence of a myriad of industry-strength DI frameworks (such as Guice~\cite{guice:19} and Spring~\cite{spring:19}), (b) industrial large adoption (i.e., it is easier to find developers to contribute with opinions about the catalog), (c) an existing specification aimed at DI (JSR-330)~\cite{javax:01}, and (d) the availability of a large number of open source software repositories written in Java.


The anti-patterns were derived from two criteria: First, based on the observation of the recurrence of bad characteristics of DI code elements, such as the violation of dependency inversion or inversion of control principles. These were observed in industry projects by the authors while maintaining software in practice and evolved through discussions with researchers. Second, as DI is supposed to improve structural quality of object-oriented applications, we also explored a set of DI anti-patterns that could be present in source code, harming design principles, such as GRASP \cite{larman:2004}.

From a methodological point of view, there are typically two approaches for coming up with new propositions: inductive and deductive \cite{sjoberg:08}. The inductive approach relies on observation of a phenomena to uncover a pattern (or set of patterns) that might lead to a theory. According to Lodico et al.~\cite{lodico:10}, "the researcher [, through inductive reasoning,] uses observations to build an abstraction or to describe a picture of the phenomenon that is being studied." In this context, we have observed the state of the practice while maintaining software projects that adopts DI framework in industrial settings.

On the other side, deductive approach concerns "developing a hypothesis [...] based on existing theory, and then designing a research strategy to test the hypothesis" \cite{wilson:10}. It means that existing theory is used as a basis for establishing a proposal so that evidence can be gathered based on a strategy developed to evaluate the proposal. For this matter, we relied on the mapping of violations of design principles (theory) , i.e. GRASP and SOLID, in the context of the adoption of DI in software projects. This mapping would enable us to hypothesize over possible anti-patterns (hypothesis).

Through maintaining software projects in industry, in efforts related to corrective and evolutionary software maintenance, one of the authors was able to preliminarily identify patterns in elements of the source code that violated design principles behind DI, namely, inversion of control and dependency inversion principles, and also some of the design principles presented in GRASP and SOLID, such as the open close principle. The experience maintaining software projects lasted 7 months, and 3 closed-source projects from an industrial partner were maintained in the period. All 3 projects were information systems developed in Java aimed at supporting business processes in different organizations in varied domains. Due to the disclosure agreement on exposing information about the closed-source projects maintained at the time, we cannot described further details about source code elements involved.

Rather than the inductive approach, deductive reasoning relies on theory to hypothesize about a phenomenon that might occur in real-world. In our context, we rely on the theory of GRASP and SOLID design principles to complement the proposition of the DI anti-patterns that might occur in practice. Some excerpts of reasoning over the existence of DI anti-patterns are provided as follows.

In regard to GRASP, the \textit{Creator} pattern advocates for reasoning upon which class (A) is responsible for instantiation of another class (B). A factor that drives the assignment of responsibility in the Creator pattern is the presence of dependencies in A that B needs in order to be instantiated by A. Thus, the presence of a direct container call harms the Creator pattern, once the container is a generic class provided by the DI framework in order to support instantiation of objects in scenarios where injection of elements is not possible, such as testing integration with third party libraries.

Next, GRASP introduces the \textit{Indirection} principle, which concerns the introduction of a mediator object in the context of two communicating objects. Indirection is achieved in DI by means of inversion of control and the employment of the DI container. It is worthy to note that any tentative to swipe the control of the framework, refraining the DI container from the responsibility to provide instances on run time, to mediate the communication between modules may be defined as an anti-pattern.

Besides, the annotation \textit{@Provides} found in the JSR-330 specification is responsible for letting the DI container aware that a given dependency must be provided by the method annotated. Thus, it is inferred that these methods are very cohesive, it is, it should enforce the principle of Low Coupling.

In addition, by receiving an injected element (it does not matter in which form), opening this specific code element for modification entails in the violation of the Open-closed Principle. The violation occurs because it is not guaranteed that the correctness of the program is maintained. Also, enforcing a design not oriented to abstractions causes the violation of the dependency inversion principle. Lastly, fabricating instances of objects in domain classes may incur in the violation of the Pure Fabrication principle.

\section{Candidate Catalog of DI Anti-patterns}
\label{sec:catalog}

Brown et al.~\cite{brown:98} advocates for a structural definition of a pattern through a template, because it "assures that important questions are answered about each pattern". Thus, similarly to Arnaoudova et al.~\cite{arnaoudova:13} and inspired by Suryanarayana et al.~\cite{refactoring}, we describe each of the candidate DI anti-patterns with a template. 

The template used to describe the dependency injection anti-patterns proposed in this work is summarized in Table~\ref{tab:template} and is composed by the following elements: name, rationale, potential causes, impacted quality attribute, pattern of occurrence, suggested refactoring, practical considerations, and Identification Approach. The name is a term used to describe the problem and its consequences in a few words~\cite{gamma:95}. The rationale explains the problem, including the context on which it is applied. The potential causes discuss the possible reasons behind the existence of the anti-pattern. The impacted quality attribute concerns the observed drawbacks of the anti-pattern in terms of software quality attributes. The pattern of occurrence exhibits and describes the structure of the anti-pattern through a source code snippet. Next, a candidate solution, by means of a suggested refactoring, describes the process on which the anti-pattern is removed and also presents a source code snippet that illustrates the snippet without the anti-pattern. Besides, we also discuss the practical considerations that should be taken into account when analyzing the anti-pattern occurrence context. Lastly, we explain the reasoning (either deductive or inductive) behind the origin of each proposed anti-pattern.

For describing the occurrence of the proposed anti-patterns, we rely on source code snippets that are based on real-world examples but adapted to ease developers' comprehension. 
One may argue that presenting source code snippets extracted directly from software projects would be more beneficial. However, our assumption for adapting was that specific project related information without the proper context would cause confusion. In fact, our experience validating the catalog confirmed our assumption. Further details on our validation are provided in Section \ref{sec:usefulness}.



\begin{table}
  \caption{Dependency injection anti-pattern template}
  \begin{tabularx}{\linewidth}{cX}
    \toprule
    \textbf{Term} & \textbf{Description}\\
    \midrule
    Name & A term that aims to express the nature of the problem concisely. \\
 \midrule
    Rationale & The justification and context on which the anti-pattern takes place \\
    \midrule
  Potential Causes & Potential causes that drive the occurrence of the anti-pattern  \\
   \midrule
  Impacted Quality Attribute & Software quality attributes impacted by the occurrence of the anti-pattern \\
   \midrule
    Pattern of Occurrence & An example highlighting the Pattern of Occurrence of the anti-pattern. \\
        \midrule
    Suggested Refactoring & A candidate solution together with its practical considerations. \\
      \midrule
    Practical Considerations & The trade offs on resorting to fixing the occurrence of the anti-pattern \\
    \midrule
    Identification Approach & The reasoning behind to come up with the proposed anti-pattern \\
  \bottomrule
\end{tabularx}
\label{tab:template}
\end{table}

In total, our candidate catalog contains twelve proposed DI anti-patterns, which are summarized in Table \ref{tab:catalogue}. In subsections ahead, we describe each DI anti-pattern. 

\newcolumntype{s}{>{\hsize=.36\hsize}X}

\begin{table*}
  \caption{Catalog of Java DI Anti-Patterns}
  \begin{tabularx}{\linewidth}{csX}
    \hline
    \textbf{Identifier} & \textbf{Name} & \textbf{Description}  \\
    \hline
    IIJ & Intransigent injection & A dependence that is not strictly necessary on construction time. However, it is provided by the DI container, thus introducing additional workload and memory consumption \\
 \hline
    CCI & Concrete class injection & Reference on a concrete class for injection rather than an interface \\
 \hline
    CPM & Complex producer method & Method that performs activities that are out of the scope of providing a dependence, often through a complex and incohesive source code \\
 \hline
    FDC & Fat DI class & Provision of a high number of dependencies to a class, hindering the modularity \\
 \hline
    USI & Useless injection & A dependence requested via DI that is not used \\
 \hline
    SDP & Static dependence provider & Usage of static fabrics or the Service Locator pattern to obtain a dependence \\
 \hline
    DCC & Direct container call & Relying directly on the DI container to obtain a dependence, introducing a direct coupling to the DI framework \\
 \hline
    OWI & Open window injection & An injected instance is passed as parameter to another class method or opened for external accessing (e.g., get method) \\
 \hline
    FCO & Framework coupling &  Elements on source code that are dependent on a given DI framework implementation \\
 \hline
    ODI & Open door injection &  An injection request is fulfilled by a DI container, however, the instance is opened for modification by an external code element (e.g., set method) \\
 \hline
    MAI & Multiple assigned injection & An injected instance is assigned to multiple attributes (may include external classes' attributes) \\
 \hline
    MFI & Multiple forms of injection &  Refers to the use of multiple forms of injection to a given element, such as through an attribute and constructor forms of injection \\
  \hline
\end{tabularx}
\label{tab:catalogue}
\end{table*}

\subsection{Intransigent Injection}
\label{subsec:intransigent}

\noindent\textbf{Rationale.} Constructor injection is a form of dependency injection where some or all of the dependencies of a given class are specified to be provisioned at construction time. In other words, during the class initialization, the DI container decidedly provides the injections prescribed by the developer. However, arbitrarily provisioning dependencies might entail undesired outcome. For instance, consider dependencies that are not strictly necessary on construction time, but rather decidedly receives an injection, introducing additional overload to the DI container.

Therefore, the anti-pattern \textit{intransigent injection} concerns dependencies that are not strictly necessary on construction time, however, they are decidedly provided by the DI container on construction time.

In summary, this anti-pattern may primarily fall on the performance as an impacted quality attribute, since it impacts the time to load components (given the additional overload on provisioning unnecessary injections) of the system.

\noindent\textbf{Potential Causes.}

\textit{Copy-paste programming.} In general developers tend to imitate existing encodings without reasoning about the potential implications of such design choices. This may be the case for the occurrence of this anti-pattern. Determining the root cause of performance problems might be complex and error-prone depending on the number of dependencies entailed by the system. A proactive behavior on specifying how dependencies are provisioned might benefit the system maintenance in the long term.

\textit{Ad hoc maintenance.} Without a strict monitoring of how the system evolves and subsequently how classes become increasingly dependent upon other components (thus impacting on performance) over time, this anti-pattern can naturally arise due to a history of system changes.

\textit{Easier testability.} Often developers invest less effort on object oriented design to achieve faster development cycles related to building unit tests. This may the case for this anti-pattern, where a developer may inject all dependencies required by the class to allow for faster development of tests.

\noindent\textbf{Impacted Quality Attributes.}

\textit{Performance}. This implementation introduces additional workload and memory consumption on construction time. In other words, overuse of object allocation in memory during construction time without the guarantee such an injection would be used in the short term. A worse scenario is observed if it's not a lightweight object, impacting on performance, since an injection that is not going to be used requires additional time in the dependency provision process given the heavy weight of such an object. Thus, this anti-pattern is categorized as a performance problem.

\textit{Understandability}. the eager injection mechanism is a factor that may lead to incorrect assumptions about the dependence provision process. For instance, less experienced developers can get confused about whether a dependence is necessary or not at construction time. 

\noindent\textbf{Pattern of Occurrence.} Figure~\ref{pic:intransigent_injection} presents the structure of occurrence, on which, although not used at construction time, the dependence \textit{example1} is provided by the DI container. Another example concerns dependencies that are unlikely to be used and, when used, are unlikely to be used again in the near future. This case also configures an intransigent injection.


\noindent\textbf{Suggested Refactoring.} A lower probability of the object usage opens up the opportunity for lazy loading depending on performance requirements of the software system. The example resolution could involve a \textit{Provider} interface, such as the one defined by \textit{JSR-330} for which similar proposals can also be found for other platforms. Such interface, when implemented is responsible for providing a given instance, solely when it is requested, rather than arbitrarily at construction time. Similar interfaces are also found in other object-oriented programming languages and frameworks. Thus, in the example resolution, an instance for the injected attribute \textit{example1} is only provided when its use is required.

\noindent\textbf{Practical Considerations.}

\textit{Avoiding Complex Code.} It is often the case the source code of a software system exhibits a widespread usage of constructor injections. Given this observation, it may be the case that such approach is preferred by software architects in order to avoid less experienced developers to misunderstand the lazy injection mechanism even though incurring in a performance and understandability trade off.

\textit{Cost-benefit Analysis.} It is worthy to consider whether such lazy injection, as a suggested refactoring, entails a benefit that pays off the code transformation effort. In some cases, even though the class injection introduces an additional overload, the refactoring may entail a higher maintenance cost and decrease software comprehension, which may not pay off in the long term.

\noindent\textbf{Identification Approach.} Inductive. Maintaining software projects on industry settings usually lead practitioners to go over debugging sessions to investigate a problem, solve a bug, or further understand the source code in the context of introducing a new feature. In this context, the authors have observed that several useless dependency provisions slowed down the maintenance process and, in the same line of reasoning, the authors have realized that such overload of dependencies at construction time could lead to a performance impact in software systems deployed in real-world scenarios.

\begin{figure}
\caption{Intransigent Injection}
\centering
\begin{lstlisting}
public class A {
	@Inject
	private IExampleInterface0 example0;
	@Inject
	private IExampleInterface1 example1;
	
	public A() { 
	    example0.doSomething();
	}
	
	public void foo() { /* omitted code */ }
	
	public void bar() {
	    example1.doSomething();
	    /* omitted code */ 
	}
}
---------------------------------------------------------------
public class A_Without_Intransigent_Injection {
	@Inject
	private IExampleInterface0 example0;
	@Inject
	private Provider<IExampleInterface1> example1Provider;
	
	public A() { 
	    example0.doSomething();
	}
	
	public void foo() { /* omitted code */ }
	
	public void bar() {
	    IExampleInterface1 example1 = example1Provider.get();
	    example1.doSomething();
	    /* omitted code */ 
	}
}
\end{lstlisting}
\label{pic:intransigent_injection}
\end{figure}

\subsection{Concrete Class Injection}
\label{subsec:concrete_class_injection}

\noindent\textbf{Rationale.} A concrete class is an actual class that can be instantiated as long as all dependencies are provided as defined by the constructor. The dependency inversion principle states that rather than concrete classes, a system should follow an interface-oriented design in order to decouple components. In this sense, a \textit{concrete class injection} concerns a dependence requested via dependency injection on which the element type (i.e., the attribute receiving the injected instance) of the dependence is a concrete class. 
This anti-pattern produces the following negative consequences. First, this solution yields a violation of inversion of control principle, once the class requesting its dependence acknowledges an implementation detail, i.e. the concrete class; second, this solution introduces less flexibility on testing, once a mock object would need to be an inherited class of the given concrete class in order to modify desired behavior; finally, according to Gamma et al.~\cite{gamma:95}, coupling to a concrete class can increase maintenance efforts.


\noindent\textbf{Potential Causes.}

\textit{Copy-paste programming.} Again, it may be the case developers simply follow a established pattern found in the source code and end up resorting to concrete class injection for evolving the software project. For instance, this was the case in several projects maintained by the authors during industrial experiences.

\textit{External dependencies.} In some cases, by relying on a framework, the developer is forced to depend upon a concrete class dependence that is established by the framework itself as part of the framework functionality. In this case, one can resort to an adapter~\cite{gamma:95}, a design pattern that encapsulates such a dependence without making the source code substantially coupled to the framework. In this case, when the framework API changes, the adapter would be the only component to adapt to the change, leaving other components of the system untouchable.

\textit{Quick-fix solutions.} Short development cycles may force practitioners to resort to fast prototyping. Developers may find it faster to simply rely on a concrete class injection even though harming design principles behind DI.

\noindent\textbf{Impacted Quality Attributes.}

\textit{Changeability and Extensibility.} Since the declared provisioned code element is concrete, the clients of the abstraction are required to depend directly on the implementation details of the abstraction.

\textit{Reusability.} With the dependency inversion principle harmed, it is difficult to reuse the abstraction in a different context other than the initial design reasoning.

\textit{Testability.} In the same line of reasoning of reusability, it is often the case where testing procedures rely on mocking objects. However, developers may face additional challenges on resorting to concrete classes when it becomes necessary to mock class' instances to perform software tests. 

\noindent\textbf{Pattern of Occurrence.} Figure~\ref{pic:concrete_class_injection} presents the structure of occurrence, where the concrete class \textit{ConcreteExample} is used for an instance injection.

\noindent\textbf{Suggested Refactoring.}
Gamma et al. \cite{gamma:95} advocates for programming to an interface, which is a natural solution to this anti-pattern. The example solution concerns following an interface oriented design when it comes to request a dependence. Further, the resolution example depicts a code transformation, on which an interface (see \textit{IExampleInterface}) is created so that the class \textit{ConcreteExample} implements it. Then, rather than relying on a concrete class injection (which configures a high coupling to class \textit{ConcreteExample}), the class \textit{B\_Without\_Concrete} now follows dependency inversion principle, once it depends on an interface (\textit{IExampleInterface}).

\noindent\textbf{Practical Considerations.}

\textit{Avoiding Unnecessary Complexity.} It may the case where the software project scope is small or non-critical (e.g., simply used as a prototype to validate ideas) and does not require a high maintenance efforts. In these cases, the use of concrete class injections may not entail an increased maintenance effort in the long term.

\noindent\textbf{Identification Approach.} Inductive. In the same line of reasoning from the last anti-pattern presented, the authors have also observed, while maintaining software systems in industry settings, a substantial amount of concrete classes being injected, which contrasts with the dependency inversion principle behind DI. In regard to projects with a reasonable amount of source code, such anti-pattern was observed to lead to increased maintenance efforts, specially in cases where some generalizability was necessary to evolve the source code.

\begin{figure}
\caption{Concrete Class Injection}
\centering
\begin{lstlisting}
public class B {
	@Inject
	private ConcreteClassExample example;
	
	private void foo(){
		example.doSomething();
		// code omitted for brevity
	}
}
---------------------------------------------------------------
public class ConcreteClassExample 
    implements IExampleInterface {

	@Override
	public void doSomething() {
		// code omitted for brevity
	}
}

public class B_Without_Concrete {
	@Inject
	IExampleInterface example;
	
	private void foo(){
		example.doSomething();
		// code omitted for brevity
	}
}
\end{lstlisting}
\label{pic:concrete_class_injection}
\end{figure}

\subsection{Complex Producer Method}

\noindent\textbf{Rationale.} A complex producer method concerns a method that performs activities that are out of the scope of providing a dependence, which must be its main objective, often through a long and incohesive code design. 

\noindent\textbf{Potential Causes.}

\textit{External Dependencies}. Besides framework-related dependencies mentioned in \textit{Concrete Class Injection}, in some cases, a dependence may require a set of business rules for instantiation including but not limited to relying on data provided by external applications, such as database systems.

\textit{Complex Business Rules}. Besides, it may be the case where a dependence necessarily requires a set of complex business rules for instantiating an object.

\noindent\textbf{Impacted Quality Attributes.}

\textit{Modularity.} A negative consequence entailed is undermining the ability of the software to adapt to change when requirements change. 

\textit{Understandability}. Resorting to a complex method to produce a given instance at runtime might negatively affect code comprehension.

\noindent\textbf{Pattern of Occurrence.} The example problem in Figure~\ref{pic:long_producer_method} shows a high complex method that should be simple, once the main concern of a \textit{Producer} method (see \textit{@Produces} annotation) is to a provide a given dependency. The DI container, when it identifies the existence of a \textit{Producer} method for a given type, transfer the responsibility for dependence provision to the Provider method.

\noindent\textbf{Suggested Refactoring.} On the other hand, in the solution part, in case where business logic is necessary to obtain a dependence, rather than relying on a \textit{Producer} method, a business method is desirable. In other words, the \textit{@Provision} annotation is removed, so the dependence provision process of the class holding the old Provision method is shortened. Also, refactoring the method in order to decrease cyclomatic complexity is another important step. In addition, a suitable code transformation is employing aspect-oriented programming in order to trigger important tasks based on the life-cycle of the Provider method. For instance, an example code transformation is defining a pointcut on the \textit{Provider} method, so the Provider method is intercepted and the logic is executed prior or after the dependence provision. Particularly, we aimed to provide an excerpt of a \textit{Producer} method without high cyclomatic complexity and fewer responsibilities.

\noindent\textbf{Practical Considerations.}

\textit{Cost-benefit analysis.} In cases where such dependence provision process necessarily requires a complex process (by either external dependencies or complex business rules) or legacy systems where the internal libraries dependencies are not well understood by the developer, the refactoring may not pay off and is a subject of analysis by the developer.

\noindent\textbf{Identification Approach.} Inductive. By investigating the source code of industry projects, the authors have observed that few dependency injections necessarily required a more complex instantiation process. For instance, dependencies on which data from a database or static files are necessary to decide for which class type will be provisioned might introduce a non-trivial source code logic. In this sense, although not enjoying popularity, this anti-pattern was born expecting that some open-source projects would show the same behavior.

\begin{figure}
\caption{Complex producer method}
\centering
\begin{lstlisting}
public class C {
    // omitted code
    @Produces
    public ProducedBean generateReport(){
        Set<Integer> selectedBacklogIds = this.getSelectedBacklogs();
        if(selectedBacklogIds == null) {
            Collection<Product> products = new ArrayList<Product>();
            productBusiness.storeAllTimeSheets(products);
            for (Product product: products) {
                selectedBacklogIds.add(product.getId());
            }
            return Action.PROCESS;
        }        
        // omitted code
        Workbook wb = this.timesheetExportBusiness.
            generateTimesheet(this, selectedBacklogIds, startDate, endDate, timeZone, userIds);
        this.exportableReport = new ByteArrayOutputStream();
        try {
            wb.write(this.exportableReport);
        } catch (IOException e) {
            return Action.ERROR;
        }
        return Action.SUCCESS;
    }
}
---------------------------------------------------
public class C_Without_Long_Producer {
    // omitted code
    @Produces
    public ProducedBean generateReport(){
        if(selectedBacklogIds == null) {
            processSelectedBacklogs();
            return Action.PROCESS;
        }        
        if (selectedBacklogIds.contains(0)) {
            processSelectedBacklogIds();
        }
        writeToLog();
        return Action.SUCCESS;
    }
}
\end{lstlisting}
\label{pic:long_producer_method}
\end{figure}

\subsection{Fat DI Class}



\noindent\textbf{Rationale.} A Fat DI class anti-pattern concerns the injection of a substantial number of dependencies in a class. This anti-pattern primarily concerns injected instances that are often inconsequentially introduced by developers without reasoning about the increased dependence of the class with other system modules. Such a pattern is often seen in projects following a layered approach where classes responsible for the business logic of the system are separated from data classes, controllers, and views~\cite{martin:96}. In this case, business classes, depending on the amount of application logic and cross-cutting responsibilities carried out, can suffer from higher coupling to other business classes, even though the dependencies are provided via dependency injection.

\noindent\textbf{Potential Causes.}

\textit{Arbitrary Evolutionary Maintenance.} Maintenance is the process of adapting the software to new demands. These demands can be in the form of bug/performance fixes or coping with new requirements, which may entail introducing additional features. In the latter, it is often the case developers arbitrarily introduce dependencies in a class. Some may think introduction of dependencies via DI does not entail increased coupling, however, even though a project follows an interface-oriented approach, the higher number of dependencies may impact the system's ability to adapt to future demands~\cite{larman:2004}.

\textit{Lack of Refactoring.}
Built from the last point, evolutionary maintenance often involves structural changes in the source code. It may be the case that, in this context, the lack of appropriate refactoring for the objective of keeping classes lower coupled can cause additional efforts on future maintenance tasks.

\textit{Inadequate Design Analysis.}
There may also be the case where an initial design analysis phase has been neglected by the team, generating a friction between the requirements pursued and design quality. It is often the case such problem occurs due to tight deadlines or resource constraints, such as limited number of developers. One may argue that this also applies to general design anti-patterns, however, this becomes more pressing in DI since the design principles behind DI provide a core rules of thumb that, if not followed, refrain the system from benefiting from DI. In sum, a negative consequence entailed by this anti-pattern concerns a possible increased effort on maintenance tasks in the class.


\noindent\textbf{Impacted Quality Attributes.}

\textit{Understandability.} The increased number of dependencies provided via DI scattered across multiple methods in a \textit{Fat DI Class} may compromise the ability of new incomers (and sometimes even expert developers) to assess what is ought to be changed in the context of corrective or evolutionary maintenance.

\textit{Changeability and Extensibility.} In the same line of reasoning of understandability, this anti-pattern can make it difficult to introduce new features or changes to the source code. This is particularly characterized when a single new feature requires changes to be performed over multiple places of the source code.

\textit{Reusability and Testability.} Reusing a Fat DI Class may be compromised given the increased number of dependencies that are forcibly required for instantiating such a class. Besides, testability can also be impacted due to the same reason.

\noindent\textbf{Pattern of Occurrence.} Figure~\ref{pic:god_di_class} depicts the Pattern of Occurrence of a God DI class. The example depicts an excerpt of a class with high level of complexity, in terms of number of injected element instances, and number of methods.

\noindent\textbf{Suggested Refactoring.} Besides, Figure~\ref{pic:god_di_class} also exhibits a suggestion of a refactoring that removes the anti-pattern, dividing dependencies and behavior into different classes. The resolution example depicts a code transformation applied to previous class D, on which a refactoring type called "Extract Class" \cite{cedrim:18} was employed three times in order to reduce the complexity of class D.

\noindent\textbf{Practical Considerations.}

\textit{Cost-benefit Analysis.} It is worthy to consider whether an extensive refactoring of the code base pays off. The reasoning is that substantial changes to a stable code base may entail the emergence of unwanted bugs. Besides, in cases where logic is scattered through the source code, such refactoring may also spread over several classes and significantly impact the effort.

\textit{Avoid Over Engineering.} Furthermore, it may be also the case where a class necessarily carries out substantial application logic. In such a cases it is worthy to consider the necessary amount of refactoring that delivers the intended outcome (e.g., a decreased degree of coupling) with some trade-off (e.g., some degree of coupling to classes that carries out correlated business logic).

\noindent\textbf{Identification Approach.} Inductive. Again, the experience of maintaining software projects allowed the authors to observe that some classes, specially those entailed for the core business logic of software systems, were carrying out a substantial amount of business logic and relying on a increased level of external dependencies (through dependency injection) to fulfill a business requirement. That necessarily leads to increased coupling among software components and then the anti-pattern was naturally induced from this observation.

\begin{figure}
\caption{Fat DI Class}
\centering
\begin{lstlisting}
public class D {
	@Inject private IExample1 one;
	@Inject private IExample2 two;
	@Inject private IExample3 three;
	@Inject private IExample4 four;
	@Inject private IExample5 five;
   // other several dependencies injected
	@Inject private IExampleN n;
	
	void methodOne() { /* reference to several dependencies */ }
	void methodTwo() { /* reference to several dependencies */ }
	// other several methods
	void methodThree() { /* reference to several dependencies */ }
}
---------------------------------------------------------------
public class D_Part_1 {
	@Inject private IExample1 one;
	@Inject private IExample2 two;
	@Inject private IExample3 three;
	
	void methodOne() { /* code omitted */ }
}
public class D_Part_2 {
	@Inject private D_Part_1 dPartOne;
	@Inject private IExample4 four;
	@Inject private IExample5 five;
	@Inject private IExample6 six;
		
	void methodTwo() { /* code omitted */ }
}
public class D_Part_3 {
	@Inject private D_Part_2 dPartTwo;
	@Inject private IExample7 seven;
	@Inject private IExample8 eight;
	@Inject private IExample9 nine;
	@Inject private IExampleN n;

	void methodThree() { /* code omitted */ }
}
\end{lstlisting}
\label{pic:god_di_class}
\end{figure}

\subsection{Useless Injection}

\noindent\textbf{Rationale.} A useless injection anti-pattern regards a dependence requested via dependency injection that is actually not used in the class requesting it.

\noindent\textbf{Potential Causes.}

\textit{Refactoring Residue}. By analyzing the root cause of this anti-pattern from both closed and open source projects, it was possible to notice that some useless injections were caused by removal of functionalities from a class that used such injections. It is worthy to note further investigation is necessary to fully understand this trend in source code.

\textit{Speculative Design}. Developers may be tempted to introduce dependencies that might be necessary in the future, which can lead to this anti-pattern.

\textit{Fear of Breaking Code}. Lastly, during the process of software maintenance, some functionalities may be removed. In such a case, whenever an injection that enabled the removed functionality is not removed, it will lead to an useless injection. However, developers may fear that, by removing such useless injection, will incur in breaking the code base.


\noindent\textbf{Impacted Quality Attributes.}

\textit{Performance}. The presence of this anti-pattern overloads the DI container with the incumbency to provide the non used dependency whenever the dependence is requested. The scenario is even worse if it is not a lightweight object or if it is not a singleton scope, impacting on performance.

\textit{Understandability.} The lack of mastery of the dependency injection technique may lead less experienced developers to refrain from removing such an injection due to the fear that it may cause the software system or some functionality to break.

\noindent\textbf{Pattern of Occurrence.}
Figure~\ref{pic:non_injection_used} presents the structure of occurrence. The example shows a class (\textit{E}) with an injected instance that is not used by any method of the class.


\noindent\textbf{Suggested Refactoring.} The solution simply concerns removing the non used injection element.

\noindent\textbf{Practical Considerations.} Unless an useless dependence is used by an inherited class, there is no practical consideration that motivates not removing such an anti-pattern.

\noindent\textbf{Identification Approach.} Inductive. The authors have also witnessed the appearance of injections that were not used during maintenance tasks. By looking at historical commits, the authors observed that most instances of useless injections were introduced in the context of a refactoring. In other words, the dependence were used at some point and, after a code refactoring, such dependence stopped being used and the developer that was performing the refactoring process has not noticed such a problem. The same pattern was also observed in open-source projects. Additional investigation is necessary to further understand such phenomena and it is out of the scope of this work.

\begin{figure}
\caption{Non Used Injection}
\centering
\begin{lstlisting}
public class E {
	@Inject
	private ExampleType one;
	
	public void foo() { /* no reference to one */ }
	public void bar() { /* no reference to one */ }
}
---------------------------------------------------------------
public class E_Without_Non_Used {
	public void foo() {
	    // code omitted for brevity
	}
	public void bar() {
	    // code omitted for brevity
	}
}
\end{lstlisting}
\label{pic:non_injection_used}
\end{figure}

\subsection{Static Dependence Provider}
\label{subsec:static_dependence_provider}

\noindent\textbf{Rationale.} Static dependence providers are associated to two forms of creating and/or providing object instances in object-oriented software systems: Fabrics~\cite{larman:2004} and Service Locators~\cite{deursen:18}. A static dependence provider refers to a static class that has the objective to provide a requested concrete implementation, not being a \textit{Provider} class. In the same line of reasoning, \textit{Service Locator} pattern also applies to the latter mentioned context, but rather to a more broader context, since it is a class that has the responsibility for serving all dependencies that might be required at run time.

\noindent\textbf{Potential Causes.}

\textit{Misconception about the Dependency Injection mechanism}. It may the case that some developers, by not understanding the role of dependency injection, fall prey on resorting to patterns that would only be necessary in the absence of dependency injection support in a software system, which is the case of a static dependence provider.

\textit{Impossibility of Applying Dependency Injection.} In cases where the software system relies on external libraries enabling some functionality, it may become necessary to explicitly handle dependency provision because the DI container may play no role outside the scope of your software system.


\noindent\textbf{Impacted Quality Attribute.}

\textit{Modularity.} A negative consequence entailed by this anti-pattern is a high coupling on fabric classes in the code base. 
In case of \textit{Service Locator}, the dependence on this pattern is even worse due to its widespread usage in the project. Indeed, inversion of control is not achieved in both cases. Both classes of problem concerns design violations, since both violate dependency inversion principle and inversion of control principle.

\noindent\textbf{Pattern of Occurrence.} Figure~\ref{pic:static_dependence_provider} depicts an example of Service Locator occurrence. 
The occurrence example exhibits the class (E) with a dependence provision performed by \textit{ServiceLocator} . In other words, rather than relying on the DI container for injecting an instance of \textit{IDataSource} type on \textit{dataSource} attribute, the code relies on the service locator pattern.

\noindent\textbf{Suggested Refactoring.} The example resolution on Figure~\ref{pic:static_dependence_provider} enforces the use of DI container for dependency injection at run time by relying on a \textit{Producer} method in order to provide an instance of \textit{IDataSource}. Particularly, the resolution example above shows a code transformation, in which the logic for creating an instance of \textit{IDataSource} is modularized within a Producer method. This way, the class \textit{E\_Without\_Service\_Locator} is not coupled to a service locator class anymore.


\noindent\textbf{Practical Considerations.}

\textit{Resource Constraints.} It is often the case software projects goes over periods where deliveries must be made with strict deadlines. In this case, introducing a technical debt regarding eschewing off dependency injection oriented code in favor of fast iterations through service locators may be a compelling strategy to cope with the deadline. However, keeping track of such technical debt is important in order to not disrupt future iterations.

\textit{Avoid Over Engineering.} Furthermore, it may be also the case where a dependence, in order to be instantiated, necessarily requires a complex dependence resolving scheme. In cases where resorting to \textit{Providers} would increase the effort on providing such an instance, a static dependence provider may be worthy. However, it is important to, in every iteration, reason about such trade-offs in the long term.

\textit{External dependencies.} In a component-oriented design, some components of the same system might not enjoy or inherit the support of dependency injection for reasons related to cross-libraries incompatibility or performance. In this case, developers might resort to static dependence providers to support dependence provision in such components.

\noindent\textbf{Identification Approach.} Inductive/Deductive. By observing direct container calls in some classes implemented in industrial projects, the authors deducted that the lack of DI knowledge could lead inexperienced software maintainers to resort to static dependence providers in software systems that enjoys DI support.

\begin{figure}
\caption{Static Dependence Provider}
\centering
\begin{lstlisting}
public class E {
	 @Inject 
	 private Parser parser;
	
    public void execute(List<String> files) throws Exception {
		IDataSourcedataSource dataSource = (IDataSource) ServiceLocator.getInstance()
		    .getBeanInstance("IDataSource");

		for(String file : files){
            Object parsedObject = parser.parse(file);
            dataSource.insert( key, parsedObject );
		}
    }
}
---------------------------------------------------------------
public class ProjectConfigBeans {
	@Bean
	public IDataSource provideDataSource(){
		// logic for creating an instance of IDataSource
	}
}

public class E_Without_Service_Locator {
	@Inject
	private Parser parser;
	@Inject
	private IDataSource dataSource;

    public void execute(List<String> files) throws Exception {
		for(String file : files){
            Object parsedObject = parser.parse(file);
            dataSource.insert( key, parsedObject );
		}
    }
}
\end{lstlisting}
\label{pic:static_dependence_provider}
\end{figure}

\subsection{Direct Container Call}
\label{subsec:direct_container_call}

\noindent\textbf{Rationale.} A direct container call can provide a concrete implementation (i.e., a class instance) at any point of the system's execution by resorting to a framework-specific component ($\S$~\ref{subsec:dependency_injection}). By resorting to a container call, the software system is highly coupled to a concrete implementation of a DI framework, which contrasts with the dependency inversion and inversion of control principles behind DI.

\noindent\textbf{Potential Causes.}

The potential causes for this anti-pattern are similar to the ones described for the \textit{Static Dependence Provider} anti-pattern ($\S$~\ref{subsec:static_dependence_provider}).



\noindent\textbf{Impacted Quality Attribute.}

\textit{Modularity}. Negative consequences include high coupling to framework specifics, since it relies directly on the framework to provide the dependency. In this sense, inversion of control principle is not achieved in this context. It is worthy to note that the nature and outcome of this anti-pattern are similar to using a static fabric or a \textit{Service Locator}. Therefore, considering that DI is chosen as an architectural standard for the project, employing direct container calls for dependence resolution conveys an architectural violation. A suggested resolution relies on applying DI on occurrences of container calls aimed at providing a dependence. 


\noindent\textbf{Pattern of Occurrence.} The Figure~\ref{pic:container_call} shows an example of container call in the Spring framework. The occurrence example shows the class (E) with a dependence provision made by a direct container call. In other words, rather than relying on the DI container for injecting an instance of \textit{IDataSource} type on \textit{dataSource} attribute, the code relies on a direct container call.

\noindent\textbf{Suggested Refactoring.} The suggested refactoring concerns removing the element that performs a container call and enforcing the use of a DI container for dependence provision. This may take place through a \textit{Provider} class, in case the dependence provision entails a more complex process, or through any form of dependency, such as attribute or constructor injection. In the Figure~\ref{pic:container_call}, an attribute injection is shown as example.

\noindent\textbf{Practical Considerations.}

\textit{Resource constraints.} Resorting to direct container calls might accelerate deliveries. However, this might come with a higher cost on software maintenance in the long term given the increased technical debt to be paid.

\textit{External dependencies.} A developer may resort to direct container calls in specific cases related to the impossibility of providing direct dependency injection support in some internal components of a system.

\noindent\textbf{Identification Approach.} Inductive. The authors have observed some instances of direct container calls in industry settings, i.e., the reliance on a strong coupling to a DI framework to resolve dependencies on the source code.

\begin{figure}
\caption{Direct Container Call}
\centering
\begin{lstlisting}
public class F {
    @Inject
    private Parser parser;
    @Inject
    private ApplicationContext context;

    protected IDataSource getRepository() {
        return (IDataSource) context.getBean("ftpDataSource");
    }

    public void execute(List<String> files) {

        IDataSource dataSource = getRepository();

        for(String file : files){
            Object parsedObject = parser.parse(file);
            dataSource.insert( key, parsedObject );
        }
    }
}
---------------------------------------------------------------
public class F_Without_Container_Call {
    @Inject 
    private Parser parser;
    @Inject
    private IDataSource dataSource;

    public void execute(List<String> files) {
	    for(String file : files){
            Object parsedObject = parser.parse(file);
            dataSource.insert( key, parsedObject );
	    }
    }
}
\end{lstlisting}
\label{pic:container_call}
\end{figure}

\subsection{Open Window Injection}
\label{subsec:open_window}

\noindent\textbf{Rationale.} The \textit{open window injection} anti-pattern is found when an injected instance is not used, but passed as parameter to another class' instance or opened for external accessing (e.g. by get method or public/protected access modifier). This scheme clearly violates the dependency inversion and inversion of control principles behind DI. The developer, along with introducing an \textit{ad-hoc} dependence passing process, clearly indicating a design violation, is also directly managing the dependence provision process, which contrasts with the idea of delegating the dependence provision process to a DI container.

\noindent\textbf{Potential Causes.}

\textit{Misconception about the Dependency Injection Mechanism}. It may be the case that some developers, by not understanding the principles behind DI,
end up relying on \textit{ad-hoc} mechanisms for exposing injected instances. The reasoning is that, to clearly separate concerns in a software system, one module should not expose the dependencies received through DI to other classes, unless strictly necessary.

\textit{Quick-fix Solutions.} Short deadlines for corrective or evolutionary maintenance may drive the opening of injected instances to the external world.


\textit{General-purpose Abstractions.} The design choices taken by developers may play a role on deciding for the usage of this anti-pattern. For instance, depending on the design choices of the project, a subclass may require accessing an injected attribute of the superclass. The same line of reasoning applies to external dependencies. It may be the case that opening the injected instance for external components is a necessary condition to implement a functionality given the previous design choices.

\noindent\textbf{Impacted Quality Attribute.}

\textit{Modularity.} Two modularity aspects are observed. Firstly, it adds a useless intermediary element between the class that needs a given concrete implementation and the DI container. Secondly, it opens a door for external modification, which could possibly yield the introduction of undesired behavior.

\noindent\textbf{Pattern of Occurrence.} Figure~\ref{pic:open_window_injection} show an example of open window injection occurrence, on which the \textit{parser} object is passed as parameter to another method.

\noindent\textbf{Suggested Refactoring.} The resolution example depicts a code transformation where the injected element \textit{parser} is not passed as parameter to method \textit{doSomething} of the interface \textit{IExampleInterface} anymore. The concrete implementation of \textit{IExampleInterface} is now responsible for defining its dependence on an instance of \textit{Parser} type.


\noindent\textbf{Practical Considerations.}

\textit{External Dependencies.} Again, external dependencies may play a role on deciding for employing this anti-pattern. It is up to the development team to reason about the trade-offs entailed by opening an injected instance for components that are external to the application.

\noindent\textbf{Identification Approach.} Deductive. The open-closed principle advocates that a module should be open for extension, but closed for modification. For this reason, the ability of object-oriented code to expose dependencies could indeed open a window for violating the open-closed principle. It is desirable that components do not outsource the dependency provision to a third component instead of the DI container. The reasoning is that, by relying on a DI container, one can avoid an increased dependency chain that could possibly lead to a costlier maintenance process.

\begin{figure}
\caption{Open Window Injection}
\centering
\begin{lstlisting}
public class F {
    @Inject
    private Parser parser;
    @Inject
    private IExampleInterface one;
    
    public Parser getParser() {
        return parser;
    }
    public void execute(List<String> files) throws Exception {
    		for(String file : files){
                Object parsedObject = parser.parse(file);
                one.doSomethingWithParsed(
                    parser, parsedObject);
    		}
    }
}
---------------------------------------------------------------
public class F_Without_Passing {
    @Inject
    private Parser parser;
    @Inject
    private IExampleInterface one;

    public void execute(List<String> files) throws Exception {
		for(String file : files){
            Object parsedObject = parser.parse(file);
            one.doSomethingWithParsed(parsedObject);
		}
    }
}

public class ConcreteExample 
    implements IExampleInterface {
    
   @Inject
   private Parser parser;

	@Override
	public void doSomethingWithParsed(Object parsedObject) {
		// omitted code
	}
}
\end{lstlisting}
\label{pic:open_window_injection}
\end{figure}

\subsection{Framework Coupling}

\noindent\textbf{Rationale.} Framework coupling refers to elements on source code that are dependent on a given framework implementation. As the name of the anti-pattern expose, it can be represented as annotations or method calls to framework configuration classes along the source code. It is worthy to note that this anti-pattern differentiates from direct container call ($\S$~\ref{subsec:direct_container_call}) since it introduces a different way of coupling to framework specific implementations, such as code annotations. This anti-pattern tends to scatter across modules of the system, whereas \textit{direct container calls} are often primarily used to achieve a particular goal that would be difficult by a traditional form of injection.



\noindent\textbf{Potential Causes.}

\textit{Copy-paste programming.} In general developers tend to create their code based on preexisting code. This may be the case for the occurrence of this anti-pattern. Determining the standard set of annotations or the standard to be followed would help to avoid this anti-pattern.

\noindent\textbf{Impacted Quality Attributes.}

\textit{Modularity.} In the context of programming platforms that presents a specification for DI, e.g., Java, a framework specific annotation incurs in coupling the application code to the framework. 

\textit{Compatibility.} In addition, in case where compatibility is a requirement, this anti-pattern can lead to greater effort in maintenance activities, a framework change, or a framework version update.

\noindent\textbf{Pattern of Occurrence.} Figure~\ref{pic:framework_coupling} depicts a class that employs Spring framework \textit{@Autowired} annotation and, below the dashed line, the same class, now employing JSR-330 \textit{@Inject} annotation.

\noindent\textbf{Suggested Refactoring.} A suitable option for removing coupling from a given DI framework is relying on the adoption of annotations and constructs presented in the standard specification of the language. 

\noindent\textbf{Practical Considerations.}

\textit{Perceived Stability of Legacy Systems.} It is often the case that systems do not change the underlying DI framework technology. Although evolutionary maintenance activities might involve updating the framework version, changing the DI framework technology is usually rare. For instance, the Spring framework historically maintains the same set of DI annotations.

\noindent\textbf{Identification Approach.} Inductive. By investigating dependency injection in open-source projects, the authors have found that some projects were relying on framework-defined code element annotations to wire dependencies in the software system.

\begin{figure}
\caption{Framework coupling}
\centering
\begin{lstlisting}
public class J {
    @Autowired
    private Parser parser;
    @Autowired
    private IDataSource dataSource;

    public void execute(List<String> files) {
	    for(String file : files){
            Object parsedObject = parser.parse(file);
            dataSource.insert( key, parsedObject );
	    }
    }
}
---------------------------------------------------------------
public class J_Without_Framework_Coupling {
    @Inject
    private Parser parser;
    @Inject
    private IDataSource dataSource;

    public void execute(List<String> files) {
	    for(String file : files){
            Object parsedObject = parser.parse(file);
            dataSource.insert( key, parsedObject );
	    }
    }
}
\end{lstlisting}
\label{pic:framework_coupling}
\end{figure}

\subsection{Open Door Injection}
\label{subsec:ope_door_injection}

\noindent\textbf{Rationale.} Dependencies requested by a class through a dependency injection mechanism are ought to be encapsulated, not exposed to external modules of the system, in order to safeguard loose-coupling and dependency inversion concerns. The open door injection anti-pattern is applied when an inject request is fulfilled by a DI container, however, the instance requested is open for modification by an external element (i.e., an object instance).

\noindent\textbf{Potential Causes.}

\textit{Copy-paste Programming.} It is often the case that, when new attributes are created, the IDE or the developer introduce \textit{setters} and \textit{getters} for such attribute. However, it may be the case that the developer ends up relying on dependency injection for such an attribute and forgets to eliminate the respective \textit{setters} and \textit{getters}, causing \textit{Open Door Injection} and \textit{Open Window Injection}.

\textit{Inadequate Design Analysis.} Given a set of design decisions taken by the developers, this anti-pattern may occur as a result of an inadequate project design. For instance, as seen hereafter, an external object instance may require to access and modify some injected instances of another class to fulfill a given business requirement. Such a phenomena may occur given suboptimal design decisions under tight deadlines or resource constraints.

\textit{Quick-fix Solutions.} During maintenance activities, developers might be tempted to expose a given code element with the assumption that such a procedure would not harm any design principle.

\textit{Lack of awareness of what should be “hidden”.} Developers might also be tempted to expose a given code element with the assumption that such a procedure would lead to a higher flexibility in designing classes.


\noindent\textbf{Impacted Quality Attributes.}

\textit{Reusability and Extensibility}. Given the injected element instance is open for modification, incorrect assumptions about the injected element might compromise the ability of reutilizing such a class.


\textit{Understandability}. Opening up an injected element may possibly lead to incorrect assumptions about system functionality. Consider that a developer expects such a class to behave in a given manner. Also consider that, an opening injection can lead other module to introduce a difference concrete implementation at runtime, thus making the class behave differently from the expected by the developer. Therefore, \textit{open door injection} can configure a hard to follow traceability of the program execution, hindering program comprehension.

\textit{Modularity.} Opening an injected instance for modification might entail harming the principles behind DI, namely, dependency inversion and inversion of control.


\noindent\textbf{Pattern of Occurrence.} Figure~\ref{pic:open_door_injection} depicts the presence of a public set method that allows changing the injected instance of \textit{parser} in runtime. Furthermore, the example depicts a public set method ("setParser"), which allows for modification of the instance of an injected element ("parser") by an external class. It is noteworthy that a public/protected injected attribute entails in the same problem explained.

\noindent\textbf{Suggested Refactoring.} The resolution shown below the dashed line in Figure~\ref{pic:open_door_injection} is the removal of the source code element (i.e., the public set method) that enables changing injected element.

\noindent\textbf{Practical Considerations}.

\textit{Necessary Exposure of Encapsulated Attributes.} Although it is often suggested that the principle of encapsulation should be safeguarded, there may be situations where exposing an injected instance might be necessary. For instance, an external module or a framework might require accessing a specific injected attribute to fulfil a given functionality.

\noindent\textbf{Identification Approach.} Deductive. The ability of object-oriented code to open up its internal behavior and attributes to the external world might be inherently harmful to DI. Since opening up external classes to modify the nature of an injection contrasts with the open-closed principle and the sovereignity of the DI container (being the one responsible for the dependency provision), the authors expected this anti-pattern to appear in software projects.

\begin{figure}
\caption{Open Door Injection}
\centering
\begin{lstlisting}
public class H {
    @Inject 
    private Parser parser;

	public void setParser(Object parser) {
		this.parser = parser;
	}

    // code omitted for brevity

}
---------------------------------------------------------------
public class H_Without_Anti_Pattern {
    @Inject 
    private Parser parser;

    // code omitted for brevity
}
\end{lstlisting}
\label{pic:open_door_injection}
\end{figure}

\subsection{Multiple Assigned Injection}

\noindent\textbf{Rationale.} Multiple assigned injection anti-pattern occurs when the reference to an injected instance is spread among multiple attributes. This can take place through two forms: (i) a superclass' attribute receiving the same injected instance reference or a module external to the class receiving such an instance (directly, by assignment or simply through open door injection). This anti-pattern is correlated to \textit{Open door injection}, since it opens a gap for an undesirable modification of the injected object at run time.

\noindent\textbf{Potential Causes.}

\textit{Inadequate Design Analysis.} 
An inadequate project design may lead to expose an injected dependence more than necessary, thus violating the open-closed principle.



\textit{Quick-fix solutions.} It may the case that such design decision might occur due to tight deadlines or resource constraints.


\noindent\textbf{Impacted Quality Attribute.}

\textit{Modularity.} As mentioned before in this section, as a result of exposing the injected attribute, the open closed principle can be violated. The developer looses the ability to track, or control the correctness of the object instance being used by the class. This can impact existing functionalities, leading to an undesirable system behavior.

\noindent\textbf{Pattern of Occurrence.} Figure~\ref{pic:multiple_assigned_injection} provides an example of occurrence of such anti-pattern. The example depicts the assignment of an injected instance of \textit{ExampleDAO} to an attribute of a parent class ("GenericBusinessImpl").

\noindent\textbf{Suggested Refactoring.} In the case of injection instance being assigned to an attribute of superclass, a better approach would be overriding an abstract method. This way, the overridden abstract method would provide the instance injected, not incurring on reference duplication. Therefore, the resolution example shown below the dashed line on Figure~\ref{pic:multiple_assigned_injection} depicts a code transformation that removes the assignment of an injected instance to an attribute presented in a parent class. The removal makes room for an abstract method in the parent class, which still allow the reference to the original injected instance.

\noindent\textbf{Practical Considerations.}

\textit{Maintenance Trade Offs.} In legacy systems, when it comes to deal with technical debt, further reasoning over the trade off of employing corrective measures and the cost on dealing with possible impact on functionalities of a stable system are usually taken into consideration. This is also the case for this anti-pattern.

\noindent\textbf{Identification Approach.} Inductive. By investigating dependency injection in open-source projects, the authors found instances of this particular form of injection. 

\begin{figure}
\caption{Multiple Assigned Injection}
\centering
\begin{lstlisting}
class ExampleBusiness extends GenericBusinessImpl {
    private IDAOexampleDAO exampleDAO;

    @Inject
    public void setExampleDAO(ExampleDAO exampleDAO) {
        this.genericDAO = exampleDAO;
        this.exampleDAO = exampleDAO;
    }

}
---------------------------------------------------------------
abstract class GenericBusinessImpl {
    abstract IDAO getGenericDAO();
}

class ExampleBusiness extends GenericBusinessImpl{

    private IDAOexampleDAO exampleDAO;
    
    @Inject
    public void setExampleDAO(ExampleDAO exampleDAO) {
        this.exampleDAO = exampleDAO;
    }
    
    @Override
    protected IDAO getGenericDAO() {
        return this.exampleDAO;
    }

}
\end{lstlisting}
\label{pic:multiple_assigned_injection}
\end{figure}

\subsection{Multiple Forms of Injection}

\noindent\textbf{Rationale.} This anti-pattern refers to the use of multiple forms of injection for a given element. As explained in \cite{LaignerMaster20}, DI frameworks allow the specification of what injections should be provisioned through several forms. The idea of using several forms of injection for the same element naturally incurs in a bad pattern. First, it is unknown how different DI frameworks react upon such anti-pattern. To the best of our knowledge, there is not a standard defined behavior among DI frameworks. While, by identifying a duplicate injection specification, some frameworks may simply discard one of them, other frameworks may even provide the same instance twice.

\noindent\textbf{Potential Causes.}

\textit{Copy-paste Programming.} In some cases, in maintenance activities, a developer may incorrectly assume a given element is not appropriately tagged for receiving an instance from the DI container and thus end up assigning another form of injection, in addition to the one already present.


\textit{Misconception About DI Mechanism.} Misconception about the DI mechanism may play a role as a potential cause. Some developers may incorrectly assume a given element must receive multiple forms of injection for working properly, which is typically not the case in DI frameworks.


\noindent\textbf{Impacted Quality Attributes.}

\textit{Understandability}. Multiple forms of injection in a module may lead to a misunderstanding of the injection process for less experienced developers.

\textit{Reliability}. In some scenarios, such as when the DI framework carries out the work of injecting an instance twice (or more, depending on the forms of injection declared for a given element), the software may behave non-deterministically. That is, injecting different instances for the same element during runtime. That would possibly lead to incorrect assumptions about the system behavior. 

\noindent\textbf{Pattern of Occurrence.} Figure~\ref{pic:multiple_forms_injection} provides an excerpt of the occurrence of this anti-pattern, where there are two forms of injection for the same element ("exampleDAO"). The first is an attribute injection. The second is a constructor injection.

\noindent\textbf{Suggested Refactoring.} The example resolution depicts only one form of injection (constructor) for the element \textit{exampleDAO}.

\noindent\textbf{Practical Considerations.} Unless there is an explicit need for multiple forms of injection, it is unlikely that such a procedure should occur for the same code element. If there is the need to inject multiple instances of the same type, it might be a better solution to declare different code elements, so each can receive a single instance, separately. This may lead to a less convoluted code. Besides, disallowing multiple forms of injection in the system may be a worthy configuration to avoid future post-maintenance problems.

\noindent\textbf{Identification Approach.} Inductive. By investigating dependency injection in open-source projects, the authors found project using multiple forms of injection for a given element.

\begin{figure}
\caption{Multiple forms of injection}
\centering
\begin{lstlisting}
class ExampleBusiness 
            extends GenericBusinessImpl {
    
    @Inject
    private IDAOexampleDAO exampleDAO;
    
    @Inject
    public void setExampleDAO(ExampleDAO exampleDAO) {
        this.exampleDAO = exampleDAO;
    }
}
---------------------------------------------------------------
class ExampleBusiness 
            extends GenericBusinessImpl{

    private IDAOexampleDAO exampleDAO;
    
    @Inject
    public void setExampleDAO(ExampleDAO exampleDAO) {
        this.exampleDAO = exampleDAO;
    }

}
\end{lstlisting}
\label{pic:multiple_forms_injection}
\end{figure}

\subsection{Summary of Contribution}

Documented DI anti-patterns in the grey literature do not directly consider the design principles behind DI, namely, inversion of control and dependency inversion principles, and the existence of design principles that guide good object-oriented design, such as GRASP and SOLID. 


Considering this scenario, we addressed our \textbf{RQ1} by applying two approaches to derive an initial catalog of DI anti-patterns. First, based on observations of bad characteristics in source code, i.e., characteristics of implementation in source code that violates design principles, such as inversion of control and dependency inversion principle. Second, as a deductive approach, we have reasoned about a set of anti-patterns that could appear in software systems that adopt DI as a mechanism to decrease coupling. As a result, we derived an initial catalog containing 12 DI anti-patterns. 
\section{Assessing Occurrence in Practice}
\label{sec:assessing}

In Section \ref{sec:proposal}, we conjectured a set of DI anti-patterns aimed at the Java platform. Hence, it is important to understand whether the proposed DI anti-patterns represent problems that are introduced by developers in practice. Thus, Sections \ref{subsec:design} and \ref{subsec:evaluating} introduces our initial efforts towards validating the proposed catalog, describing the steps taken to develop a static analysis tool to automatically detect instances of DI anti-patterns. Next, Section \ref{sec:evaluation_results} describes the results of the detection of DI anti-patterns in several software systems. 

\subsection{Designing a Detection Tool}
\label{subsec:design}

At the time we were investigating feasible approaches to enable a fast process of identification of the proposed DI anti-pattern instances on source code, we did not find any tool that would easily allow us expressing the rules that flag elements of code as positive or negative regarding being an instance of anti-pattern.

Although no tool was able to automatize all steps per se, (namely, querying a repository of software project through a query language, cloning the repositories queried, submitting rules to identify anti-patterns in a project's source code, and formatting and outputting the results), we found that \textit{Repodriller} \cite{repodriller:19}, a framework for mining software repositories, would enable the identification of annotations in source code. Thus, we started with a manual analysis aimed at providing initial evidence that the candidate DI anti-patterns have instances on real open source projects. \textit{Repodriller} \cite{repodriller:19} was used to filter the occurrence of DI injection point annotations (such as \textit{@Inject} and \textit{@Autowired}) and DI container references (e.g. direct container calls) in the preliminary selected projects. After filtering, classes and its associations were manually analyzed in order to verify if DI elements on source code incurred in an anti-pattern instance.

Although we were able to identify some instances of anti-patterns in the source code of randomly selected projects, we realized that the manual procedure was error-prone and not time-effective. Thus, in order to support the automatic detection of each proposed DI anti-pattern in source code, a software tool called DIAnalyzer was developed. The source code of DIAnalyzer is available on GitHub \footnote{https://github.com/rnlaigner/dianalyzer} under a MIT license to allow further exploration.

Based on the proposed catalog of DI anti-patterns presented in Section \ref{sec:catalog}, this section aims at describing the development of a software system called DIAnalyzer that automatically identifies every anti-pattern proposed. The tool is a static code analyzer implemented using the JavaParser \cite{smith:18} library, which relies on an Abstract Syntax Trees (AST) in order to flag elements of code that represent DI anti-patterns candidates. Additional details and the requirements of the tool are explained hereafter.

\begin{table}
\centering
  \caption{Rules for anti-patterns detection}
\begin{tabularx}{\columnwidth}{c|X}
     \hline
    \textbf{Identifier} & \textbf{Rule}\\
     \hline 
    IIJ & IIJ is applied when an injected attribute is not referenced in all methods of a class \\
    \hline 
    CCI & CCI is applied when an attribute that receives an injection is a concrete implementation \\
    \hline 
    CPM & CPM is applied when the sum of the cyclomatic complexity of the Producer method is greater than 8 \\
    \hline 
    FDC & FDC is applied when the sum of the cyclomatic complexity of all methods of the class being inspected is greater than 46 and the number of attributes injected in class being inspected is greater or equal 5 \\
    \hline 
    USI & USI is applied in an attribute if this attribute is an injected attribute, however, the same is not used in the class \\
    \hline 
    SDP & A heuristic was used to detect these instances, as follows: An attribute instance is obtained by calling a dependence on which its name or class name contains fabric or factory \\
    \hline 
    DCC & DCC is applied when an instance of \textit{ApplicationContext} class calls the method \textit{getBean} \\
    \hline 
    OWI & OWI is applied when an injected instance is passed as parameter to another class method or opened for external accessing by a method \\
    \hline 
    FCO & FCO is applied when the annotation \textit{@Autowired} is employed in order to inject dependence instances \\
    \hline
    ODI & ODI is applied when an injected instance is allowed to be changed on a public set method \\
    \hline 
    MAI & MAI is applied when more than one class attribute receives exact same injected instance \\
    \hline 
    MFI & MFI is applied when a class attribute is registered to receive an injected instance by more than one form of injection (e.g. constructor and attribute) \\
   \hline
\end{tabularx}
\label{tab:rules}
\end{table}

\begin{figure}
\centering
\includegraphics[width=0.5\textwidth]{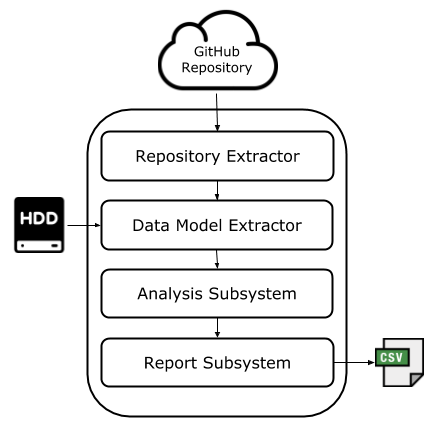}
\caption{Schematic overview of DIAnalyzer}
\label{fig:schematic_v2}
\end{figure}

Figure~\ref{fig:schematic_v2} shows a schematic overview of DIAnalyzer. The design of the project followed an orientation to abstraction, as suggested by Martin \cite{martin:96}. Thus, the system architecture is decomposed in a set of subsystems. The subsystems are: repository extractor, data model extractor, analysis, and report. The description of each subsystem is provided as follows.

The \textit{Repository Extractor} subsystem is responsible for submitting a request (in form of a query) to GitHub API in order to clone a set of open source projects into user's file system. 

As mentioned earlier, to support the identification of the DI anti-patterns, the \textit{JavaParser} framework was employed. The Data Model Extractor subsystem converts each file of the project under analysis into a model that can be manipulated by the system. The Data Model Extractor Subsystem is built as a layer above the \textit{JavaParser} framework, abstracting its internals in order to ease reuse and diminish coupling to \textit{JavaParser} from other subsystems of the architecture. \textit{JavaParser} relies on constructing an AST for a given compilation unit (i.e. Java class) and represents object oriented elements, such as methods, attributes, and classes in form of a vertex in a tree.

A rule-based strategy approach was employed to identify the DI anti-patterns. For example, to check whether a class contains the DCC anti-pattern, in the case of a project employing the Spring framework, we first identify the presence of a coupling to \textit{ApplicationContext} class. Then, based on an attribute declaration of type \textit{ApplicationContext}, we identify method calls to \textit{getBean} from this attribute, passing a string as a parameter. This string identifies either a desirable concrete class or an interface. If there are at least one method invocation of this nature, this code snippet is flagged as containing DCC. The rules applied in order to detect each DI anti-pattern are found on Table~\ref{tab:rules}. It is worth of mention that the detection strategies applied to CPM and FDC were based on Lanza and Marinescu ~\cite{lanza:95}, which are also employed by relevant studies on detection of code smells~\cite{cruzes:10}.

The \textit{Analysis} subsystem is the core part of the architecture, since it contains the logic that verifies if an anti-pattern is applied in the context of a class. Basically, each anti-pattern is modeled as a class in the Analysis subsystem. An anti-pattern class is composed by a set of rules. Each rule is also modeled as a class, being responsible for identification of injected elements that obey the characteristics of the given rule. Some rules also have as dependence a data source, which is implemented as a class that provide additional information about the source code on run time. 

Lastly, the \textit{Report} subsystem is responsible for handling requests related to convert the results of the Analysis Subsystem into a report.

\subsection{Evaluating DIAnalyzer Detection Tool}
\label{subsec:evaluating}

Although building a static analysis tool would allow us to efficiently mine software repositories, threats of validity could be risen against the effectiveness of the tool. Works that present propositions in form of catalogs (such as Chen and Jiang~\cite{chen:17}) usually rely on an oracle data-set in order to conduct an evaluation of tools that are built with the objective to flag instances present in the catalog. 

However, as there is no available oracle data-set which contains the verified instances of the DI anti-patterns we propose in this work, in order to evaluate DIAnalyzer, we built an oracle by ourselves. The first author of this paper randomly selected a set of classes from latest releases of two projects (Agilefant and Libreplan). These projects were randomly selected among projects with representative usage of JSR-330 annotations. Then, the first author manually identified 141 occurrences of DI anti-patterns (89 from Agilefant and 52 from Libreplan) related to 83 different classes (43 from Agilefant and 40 in Libreplan), concerning eight different DI anti-patterns. Thereafter, the instances identified were handed over to an independent researcher that performed a double check on the manually detected instances. Then, the independent researcher randomly selected a set of instances for each DI anti-pattern in both projects. In total, 43 manually detected instances were reviewed, confirming them as correctly identified anti-patterns. Nevertheless, although no significant disagreements were found\footnote{A misconception regarding the occurrence of one of the anti-patterns was observed and subsequently mitigated, giving us the confidence that the double check process has achieved its objective.}, we are aware that this activity is naturally error-prone and that our oracle may still miss some instances. 

We have conducted a relative recall analysis of DIAnalyzer considering the manually generated oracle. During this analysis, our tool was able to retrieve 130 out of the 141 manually identified instances, including instances of all eight DI anti-patterns contained in the oracle, resulting in a relative recall of 92.19\%. Hence, we were confident that the tool can effectively detect anti-pattern instances. A more detailed analysis on the precision identifying each DI anti-pattern follows.

In order to calculate the precision, we have manually examined every DI anti-pattern detected by DIAnalyzer in a randomly selected scope of classes (43 from Agilefant and 39 from Libreplan). Table \ref{tab:tool_evaluation} shows the precision results. Each row corresponds to an anti-pattern and each column refers to the precision. DIAnalyzer detected 835 instances of DI anti-patterns, with precision between 80 to 100\% for IIJ, CCI, FDC, USI, SDP, OWI, FCO, ODI, MAI, and MFI. The reason for the 40\% precision on Libreplan regarding DCC is due to a malformed output of the tool, which duplicates the instance found. As a consequence, several DI anti-patterns were informed more than once, harming the precision results. We have also calculated the average precision of DIAnalyzer per project. The average precision for Agilefant was 97.78\% and the average precision for Libreplan was 89.80\%. We considered these precision results to be sufficient for our purpose of evaluating the occurrence of the DI anti-patterns in Java projects.

\begin{table}
\centering
  \caption{Precision results of DIAnalyzer}
  \label{tab:tool_evaluation}
  \begin{tabular}{cccc}
    \toprule
    \multirow{2}{*}{\makecell{\textbf{DI} \\ \textbf{Anti-Pattern}}} &
    \multicolumn{2}{c}{\textbf{Project}} \\
    \cline{2-3}
    & \textbf{Agilefant} & \textbf{Libreplan}
    \\
    \midrule
    IIJ & 100\% (152/152) & 100\% (145/145) \\
    \midrule
    CCI & - (0/0) & 100\% (24/24) \\
    \midrule
    FDC & 100\% (7/7) & 80\% (4/5) \\
    \midrule
    USI & 100\% (19/19) & 90\% (18/20)  \\
    \midrule
    SDP & - (0/0) & 100\% (5/5)  \\
    \midrule
    DCC & 100\% (2/2) & 40\% (2/5)  \\
    \midrule
    OWI & 80\% (4/5) & 88\% (30/34) \\
    \midrule
    FCO & 100\% (152/152) & 100\% (144/144) \\
    \midrule
    ODI & 100\% (84/84) & 100\% (2/2) \\
    \midrule
    MAI & 100\% (25/25) & - (0/0) \\
    \midrule
    MFI & 100\% (1/1) & 100\% (4/4) \\
    \bottomrule
  \end{tabular}
\end{table}

\subsection{Detecting DI Anti-patterns}
\label{sec:auto_detecting}

With a satisfactory result in the precision and recall evaluation carried out in DIAnalyzer, we are confident that DIAnalyzer can effectively flag instances of DI anti-patterns from source code. Then, we have divided the process of detecting instances of candidate anti-patterns in software projects in two steps: open source and closed-source applications.

\subsection{Open Source Software Systems}

Mining open source software repositories constitute a common research practice in the software engineering field. For instance, studies on code smells \cite{cruzes:10} and refactoring \cite{cedrim:18} often rely on source code repositories to support their analysis. In line with this method, we found worthy to start the analysis of DI anti-pattern instances with open source repositories.


Since we have a tool that automatically identifies anti-pattern instances on source code, no manual analysis is required. Therefore, the first step is to choose a suitable set of software projects. GitHub was chosen as the repository source of software projects. Our study selected four GitHub projects that meet the following quality criteria: (i) Dependency injection usage within the project, i.e., employing a DI framework, such as the one provided by Spring; (ii) historical developer engagement with several commits; (iii) source code repository mainly written in Java. The list of selected projects is in Table \ref{tab:selected_projects}, presenting the (i) name, (ii) Java lines of code, and (iii) number of commits for each project.

\begin{table}
\centering
  \caption{Selected projects}
  \label{tab:selected_projects}
  \begin{tabular}{cccc}
    \toprule
    \textbf{Index} & \textbf{Name} & \textbf{LOC} & \textbf{Commits}\\
    \midrule  
    P1 & Agilefant & 58.171 & 5.166 \\
    P2 & BroadleafCommerce & 327.058 & 9.146 \\
    P3 & Libreplan & 284.090 & 9.659\\
    P4 & Shopizer & 109.792 & 305 \\
  \bottomrule
\end{tabular}
\end{table}


We applied DIAnalyzer on the latest releases of the four selected projects. The detection results are depicted in Table~\ref{tab:occurrence}. It is possible to observe that IIJ, CCI, FDC, USI, DCC, OWI, FCO, and ODI have instances in all four projects. Additionally, all four projects present anti-pattern instances for each DI anti-pattern category (\textit{cf.} Chapter \ref{sec:catalog}).

The large number of instances for IIJ and FCO for almost all of the analyzed projects (except for the occurrences of FCO in project P4) is noteworthy. We believe that the large number of IIJ occurrences is due to the lack of judgment by developers over the need of introducing extra injections in a class. Regarding the large number of FCO occurrences, it can be explained by a wide adoption of a Spring specific annotation \textit{@Autowired}. On the other hand, MAI only had instances in P1, suggesting a design choice that led to this anti-pattern in this specific project. CPM was not found in any project, suggesting that developers of the analyzed systems are aware that dependency provision methods must be highly cohesive and present low complexity. 

\begin{table}
\centering
  \caption{Occurrence of DI Anti-Patterns in open source projects}
  \label{tab:occurrence}
  \begin{tabular}{ccccc}
    \toprule
    \multirow{2}{*}{\makecell{\textbf{DI} \\ \textbf{Anti-Pattern}}} &
    \multicolumn{4}{c}{\textbf{Project}} \\
    \cline{2-5}
    & \textbf{P1} & \textbf{P2} & \textbf{P3} & \textbf{P4} 
    \\
    \midrule
    IIJ & 366 & 1127 & 1149 & 854 \\
    \midrule
    CCI & 3 & 277 & 52 & 185 \\
    \midrule
    CPM & 0 & 0 & 0 & 0 \\
    \midrule
    FDC & 11 & 20 & 22 & 22 \\
    \midrule
    USI & 41 & 215 & 101 & 161 \\
    \midrule
    SDP & 6 & 21 & 35 & 0 \\
    \midrule
    DCC & 4 & 45 & 20 & 3 \\
    \midrule
    OWI & 13 & 122 & 167 & 110 \\
    \midrule
    FCO & 367 & 152 & 1102 & 3 \\
    \midrule
    ODI & 114 & 90 & 2 & 15 \\
    \midrule
    MAI & 37 & 0 & 0 & 0 \\
    \midrule
    MFI & 1 & 0 & 5 & 1 \\
    \bottomrule
  \end{tabular}
\end{table}

\subsection{Closed Source Software Systems}

An important step in the software engineering field is making sure propositions reflect on the practice of software engineering. Without such validation, it is unknown whether the candidate anti-patterns are relevant in industrial settings. However, researchers often rely on mining open source repositories due to the complexities involved in obtaining closed-source software repositories. Legal issues concerning strategic processes a software supports are an example of such impedance.

At this point, we were already aware that the proposed instances of DI anti-patterns occur within popular open source software systems. However, we would like to investigate whether the proposed DI anti-patterns also occur in industrial settings. Besides, in the case of verified occurrences, we would like to comprehend if the same trends found in open source repositories are also found in closed-source repositories.

Thus, intending to obtain the source code of industrial software systems, we identified two candidate companies that the author has had previous working experience in software development and maintenance. We designed a consent term safeguarding the companies against any misuses of the source code provided. We sent them the term along May and June, 2019. The companies responded positively and expressed their willingness to support the research being conducted. Both firms provided one software for analysis. It is important to mention that the projects obtained from the firms were not developed or maintained by the author of this work.


In total, 2 projects from two different industry partners were obtained. Table~\ref{tab:closed_source_projects} shows the characteristics of both projects. In this work, we cannot expose details about the software due to non-disclosure agreement restrictions, so we cannot give full details about the closed-source projects under analysis. In short, both software projects are web-based information systems written in Java, following a layered-oriented design as found in the open source projects analyzed, and present approximately 30K LOCs. 

The first (CS1) adopts Spring \cite{spring:19}, a framework that already provides DI capabilities. CS1 is a shorter (in LOC) Java project compared to open source projects we previously analyzed. By analyzing the project, we observed and confirmed with the firm that the project was mainly developed and maintained by a senior developer, who was already experienced in development with the Java platform. The second closed-source project (CS2) employs Guice framework to support DI capabilities. Again, CS2 is shorter in LOC compared to open source projects. As confirmed with the firm, CS2 was also mainly developed by a senior developer.

Due to the use of Guice framework on CS2, we have adapted DIAnalyzer to support annotations present in Guice (e.g. \textit{@Produces}) and Guice direct container calls. In addition, we have fixed the bug reported in Section \ref{subsec:evaluating} about the malformed output of the tool in DCC. Then, we have applied DI Analyzer to automatically detect instances of candidate anti-patterns in the closed-source projects obtained. The results are shown in Table~\ref{tab:occurrence_2}.

\begin{table}
  \centering
  \caption{Selected projects}
  \label{tab:closed_source_projects}
  \begin{tabular}{ccc}
    \toprule
    \textbf{Index} & \textbf{LOC} & \textbf{Framework}\\
    \midrule  
    CS1 & 29.405 & Spring \\
    CS2 & 32.204 & Guice \\
  \bottomrule
\end{tabular}
\end{table}

\begin{table}
\centering
  \caption{Occurrence of DI Anti-Patterns in closed-source projects}
  \label{tab:occurrence_2}
  \begin{tabular}{cccc}
    \toprule
    \multirow{2}{*}{\makecell{\textbf{DI} \\ \textbf{Anti-Pattern}}} &
    \multicolumn{2}{c}{\textbf{Project}} \\
    \cline{2-3}
    & \textbf{CS1} & \textbf{CS2}
    \\
    \midrule
    IIJ & 68 & 265  \\
    \midrule
    CCI & 2 & 12  \\
    \midrule
    CPM & 0 & 0  \\
    \midrule
    FDC & 1 & 4  \\
    \midrule
    USI & 15 & 86  \\
    \midrule
    SDP & 1 & 3  \\
    \midrule
    DCC & 0 & 0  \\
    \midrule
    OWI & 64 & 35 \\
    \midrule
    FCO & 68 & 0  \\
    \midrule
    ODI & 3 & 0  \\
    \midrule
    MAI & 0 & 0  \\
    \midrule
    MFI & 0 & 0  \\
    \bottomrule
  \end{tabular}
\end{table}

Overall, the patterns observed in closed-source repositories are closely related to the findings of the previously analyzed open source projects. For instance, IIJ, CCI, FDC, USI, and OWI have instances in both projects. Regarding CPM, no instances were verified again. In line with the open source projects, both closed-source projects present anti-pattern instances for each DI anti-pattern category (\textit{cf.} Chapter \ref{sec:catalog}).

Anti-patterns IIJ, CCI, FDC, USI, and OWI have instances in both closed-source projects. Hence, although expert developers tend to follow an interface-oriented design and avoid classes that centralize the intelligence of the system, they may introduce some anti-pattern instances in source code. This can be explained by fast prototyping sprints in the software life cycle and lack of attention (in case of USI). Instances of FCO solely found in CS1 is explained by a wide adoption of the Spring annotation \textit{@Autowired}. Since CS2 employs Guice, a framework that follows JSR-330 convention of \textit{@Inject} annotation, no instances of FCO are verified in CS2.

Besides, it is observed that, in opposition to the findings in open source repositories, DCC, MAI, and MFI have no instances in CS1 and CS2. In addition, ODI are verified only three times (CS1). We believe these results are related to three main factors: (i) the LOC of the closed-source repositories, which are smaller compared to the analyzed open source projects, (ii) the number of developers involved in development activities, which is also smaller compared to the open source projects analyzed, and (iii) the expertise of the developers involved in the development of the closed-source projects. 

In other words, it may be the case that the expert developers involved in development activities of the closed-source projects analyzed are aware of the risks entailed to the software architecture by: the introduction of direct container calls (DCC); opening injected fields to external modification (ODI); the assignment of instances provided by the DI container to several fields (MAI); the introduction of multiple forms of injection for a given element (MFI).

\subsection{Discussion}
\label{sec:evaluation_results}

The characteristics of the open source and closed-source projects analyzed differ profoundly. For instance, the number of lines of code and the number of developers involved in the development process vary greatly. In open source projects, the smallest project (P1) has 58.171 LOCs and a high number of developers are involved, as investigated in the commit history. In opposite, in the closed-source projects, we were not able to gather and analyze projects with similar characteristics to the open source ones. Both closed-source projects analyzed are small (in terms of LOC) in comparison with the open source projects and were mainly developed and maintained by few expert developers each, which may be one of the reasons why some anti-patterns occurrences were not observed.

Some DI anti-patterns are prominent in both open and closed-source, such as IIJ, CCI, FDC, USI, SDP, OWI, and ODI. We believe this trend occurs due to fast development sprints (FDC, SDP), lack of knowledge about principles behind DI (CCI, SDP, DCC, ODI), and misuse of DI framework and lack of attention (IIJ, USI). In addition, DCC, ODI, MAI, and MFI are mostly observed in open source projects. We believe that lack of proper knowledge of DI hinders avoiding these instances in source code. As expected, since CS2 employs Guice, FCO did not appear in its source code. Guice framework relies solely on JSR-330 specification to define injection points and does not have annotations different from the definition as Spring does. Lastly, in closed-source projects, CPM again does not have any instance as found in open source projects. Also, DCC, MAI, and MFI does not occur in closed-source projects.

Although not conclusive, some aspects may have influenced the uncovered trends, which we address next. By reviewing the historical commits of open source projects, we verified that anti-patterns progressively scatter around source code during the development process. One of the possible reasons is that novice developers tend to base their code on previously committed code. Another characteristic found in open source DI supported projects that have a significant number of anti-patterns instances is the number of developers that actively contributed to the code base. In other words, we observed that projects with multiple developers tend to show more anti-patterns instances compared to those with less contributing developers. It is worthy to note these are hypotheses that should be investigated in more depth.


Lastly, although we were able to confirm the existence of our proposed anti-pattern instances in both popular open source and industrial software projects, it is worthy to note that the results of our assessment do not allow to draw a precise empirical profile of the DI anti-pattern occurrence, which is out of the scope of this paper.
\section{Investigating Perceived Usefulness of Proposed Catalog}
\label{sec:usefulness}

In the last Section (cf. \ref{sec:assessing}), we verified that the conjectured anti-patterns occur within software systems. Although instances of our candidate anti-patterns are retrieved from both closed and open-source software systems, an important step towards strengthening the validation of the proposed catalog of anti-patterns is gathering the perception of experienced practitioners. In other words, we aim to understand if industry practitioners consider the catalog useful and are willing to apply our catalog in their working environment.

Thus, in this chapter, we document our efforts to investigate the acceptance of our catalog among industry practitioners by the application of the Technology Acceptance Model (TAM)~\cite{davis:89} in its three dimensions: ease of use, usefulness, and intention of use. Hence, besides investigating the occurrence of each DI anti-pattern in software projects as explored in the previous chapter, we have designed and conducted an interview-administered survey and an online survey to assess the acceptance and perception of usefulness from expert developers regarding the proposed catalog. 

Obtaining the perception of experienced developers over the candidate catalog proposed is an important validation step prior to sharing it with the software engineering community. Using the GQM (Goal Question Metric) definition template described by Wohlin et al.~\cite{wohlin:12}, our goal can be further defined as: \textit{Analyze} the proposed catalog of DI anti-patterns \textit{for the purpose of} characterization \textit{with respect to} the acceptance and perceived usefulness \textit{from the point of view of} software developers with large industrial experience applying DI \textit{in the context of} software projects employing dependency injection.

As TAM is employed to assess a given technology (in our case, the candidate catalog), we want to assess the willingness of the expert developers to adopt the catalog as a tool in their development activities. Also, we aim to gather a preliminary assessment of difficulties found by developers on understanding our catalog. The rate of answers a certain facet might indicate opportunities for improving the catalog.

To achieve the goals, this study designed two classes of surveys to gather the opinion of developers over the candidate catalog. As mentioned at the beginning of this chapter, we first designed an interview-based questionnaire to allow an in-depth analysis of the instrumentation and each candidate anti-pattern proposed. Then, with the lessons learned from the interview-based survey and corrections leveraged by opinions of expert developers, an online survey was designed to obtain a wider range of views regarding the proposed catalog. Lastly, the results, lessons learned, and threats of validity are explained.

\subsection{Summary of the Methodology}

Here we summarize the multi-phase survey approach we designed to validate the catalog. We started (Section~\ref{sec:interview}) with three very experienced developers to both (i) validate the instrumentation used to assess the catalog and (ii) validate the ideas proposed in the catalog in terms of their soundness and validity. The input received from this phase, which was conducted through direct interviews, gave us confidence to resort to a larger corpus of experienced developers. 

In the second phase (Section~\ref{sec:online}), we selected many expert developers by convenience and conducted the survey through an online questionnaire. From eleven experienced developers contacted, six provided responses.

To improve the number of responses, in the final phase, we made the questionnaire openly available online to all kinds of developers (Section~\ref{sec:online}). Given the extension of the questionnaire, we divided it into two distinct questionnaires. In this phase, we received input from five additional experienced developers, but also from four novice developers. In total, this phase received 9 more responses.

It is worthy to note that, along the next section, the surveying process is described in chronological order and the multi-phase fashion was employed to incrementally validate the catalog towards adapting the questionnaire to be openly available online so we could reach out to more practitioners.

\subsection{Interview-Based Survey}
\label{sec:interview}

This section presents the details of an interview-based survey conducted in order to obtain preliminary results about the usefulness of our proposed catalog.

\subsubsection{Design}

Towards achieving our goals, we first designed a descriptive survey. According to Linaker et al. \cite{mainani:survey}, a descriptive survey supports claiming or assertions about a particular subject. Thus, as we are claiming that our candidate catalog provides a comprehensive set of DI anti-patterns, a descriptive survey meets our goal. Regarding the target population, for this preliminary study, we followed the recommendation to select developers that are most appropriate for our goal in order to provide accurate answers, rather than expecting that a random target population would allow an effective analysis of our subject \cite{wagner2020challenges}. Indeed, we targeted at a population of practitioners with large expertise on applying design principles, frameworks, and dependency injection in software systems.

An interviewer-administrated questionnaire was designed in order to avoid threats of validity, such as doubts that could arise during the process, then leading to a wrong answer by the respondent. Linaker et al. \cite{mainani:survey} assert that employing thus questionnaire type enables clarifying ambiguous questions. Thus, we aimed to further support the interviewees in comprehending the context, purpose, and consequences of each candidate anti-pattern proposed during the interview.

The questionnaire is divided in three parts, as explained as follows. The first part concerned gathering information about the respondents' academic background and industrial experience. For instance, questions included years of experience developing software and current position in industry. Regarding technical skills, we inquiry about object-oriented analysis and design, design principles and patterns, anti-patterns, source code inspection, dependency injection, and Java programming language.
In addition, we collected information about English reading and comprehension skills.

The second part of the questionnaire consisted of questions regarding the candidate catalog. For each DI anti-pattern proposed, following the pattern structure mentioned in chapter \ref{sec:proposal}, the information provided concerns: (i) the name of the DI anti-pattern, (ii) a short description of the DI anti-pattern, (iii) a characterization of occurrence in form of source code, (iv) the negative consequences, (v) a description of a possible resolution, and (vi) a characterization of resolution in form of source code.

Based on the information provided, the interviewees were inquired to answer the following question: "Can the proposed DI anti-pattern actually be characterized as an anti-pattern?" The question is responded based on a five-point Likert scale (1- Agree, 2- Partially Agree, 3- Neutral, 4- Partially Disagree, and 5-Disagree). For this specific design survey, the interviewees were also invited to include comments over the general structure of the analyzed DI anti-pattern and possible disagreements with the anti-pattern or even about the resolution example provided.

The final part of the questionnaire concerned the application of the Technology Acceptance Model (TAM) \cite{davis:89}. According to Turner et al. \cite{turner:10}, TAM is a suitable tool to capture the user's acceptance of a given technology. The technology, in our case, is the candidate catalog and the anti-patterns proposed within it. TAM questions aim to assess three acceptance model constructs: usefulness, ease of use, and intention to use. The complete instrumentation employed in our survey can be accessed online\footnote{https://zenodo.org/record/3066339}.

\subsubsection{Execution}

The execution of the survey followed the strategy of identifying a sample of the population according to the survey design. Thus, we prioritized experts in software development, those with extensive experience in industry on developing and maintaining software systems, particularly with source code inspection, design patterns, and software design and architecture skills.

Then, we identified three interviewees from three different organizational units of two different industrial partners. The identified interviewees are further described in Table~\ref{tab:background}. It is observed that the interviewees have long-lasting experience in industry and a strong background in software development, being a reliable source when it comes to evaluating the candidate catalog. It is noteworthy to mention that we intentionally opted for selecting a small sample of experts to conduct in-depth interviews, allowing qualitative discussions about our initial DI anti-patterns catalog.

Table~\ref{tab:background} presents background information of the interviewees, in which it is possible to observe that they indeed have a strong background in object-oriented analysis and design, being a suitable source when it comes to evaluating the proposed catalog. Although I3 does not possess a strong experience in DI, I3 was able to assess the proposed anti-patterns due to having strong software design skills. Due to the format of the questions and the questionnaire type, the survey was provided through a printed document. The interviewees were informed that there was no limit of time. The interviews took place in April 2019, and lasted 80, 85, and 100 minutes, respectively for interviewees I1, I2, and I3.

\begin{table*}
  \caption{Background of respondents}
  \label{tab:background}
  \begin{tabularx}{\linewidth}{XXXX}
    \toprule
    \multirow{2}{*}{\textbf{Information}} &
    \multicolumn{3}{c}{\textbf{Respondent}} \\
    \cline{2-4}
     & \textbf{I1} & \textbf{I2} & \textbf{I3} \\
    \midrule
     Academic background & Master & Bachelor & PhD \\
    \midrule
    English reading and comprehension skills & Advanced & Advanced & Advanced \\
    \midrule
    Experience developing software & > 10 years & > 10 years & > 10 years \\
    \midrule
    Current position & Project Manager & Tech Leader & Tech Leader\\
    \midrule
    Object-oriented analysis and design & Several projects in industry & Several projects in industry & Several projects in industry\\
 \midrule
    Design principles and patterns & Several projects in industry & Several projects in industry & Several projects in industry \\
    \midrule
    Anti-patterns & Several projects in industry & A project in industry & A project in industry \\
    \midrule
    Source code inspection & Several projects in industry & Several projects in industry & Several projects in industry \\
 \midrule
    Dependency injection & Several projects in industry & Several projects in industry & A project in industry \\
 \midrule
    Java & Several projects in industry & A project in industry & Several projects in industry \\
    \bottomrule
  \end{tabularx}
\end{table*}

\subsubsection{Results}

The results of the survey are presented in Tables \ref{tab:perception} and \ref{tab:tam}.

The results of the perception on the proposed DI anti-patterns are shown in Table~\ref{tab:perception}. It is noteworthy to mention that from 39 inquiries over DI anti-patterns, we observed only 2 (partial) disagreements (IIJ and FCO). IIJ is the only anti-pattern proposed that does not have any full agreement response. I1 mentions that "IIJ does not yield an anti-pattern when it comes to lightweight objects." In addition, I2 asserts that "dependencies that are not needed on construction time should be moved to another class in order to save resources." For FCO, I2 argues that "most projects do not change the chosen DI framework" and I3 argues that "the anti-pattern applies only when compatibility is defined as a requirement." 

On the other hand, 37 responses concerned "Agree" (31) and "Partially Agree" (6) responses, which yields 94.8\% of the total answers. In addition, from the 13 proposed anti-patterns, 11 contain at least two full agreements. Most of these are from the architecture and design problems categories. Regarding the partially agree responses, in CPM, I1 agreed the occurrence is bad, but argued that "it is not directly related to DI." Also, in relation to MAI, I1 did not agree with the example solution provided, arguing that "the problem exposed in the structure of occurrence is a poorly implemented refactoring." The comments provided by the respondents suggest that the partial agreements regard context-based situations (e.g., situations in which the code structure represents a problem depending on the requirements). We believe that this result reflects the fact the complete context information of the anti-patterns was not included in the survey. Overall, given the experience of the respondents on design principles and patterns, these observations provide a positive perception of the catalog.

\begin{table}
\centering
  \caption{Perception over the DI anti-patterns in interview-based survey}
  \label{tab:perception}
  \begin{tabular}{cccc}
    \toprule
    \multirow{2}{*}{\makecell{\textbf{DI} \\ \textbf{Anti-Pattern}}}
    & \multicolumn{3}{c}{\textbf{Respondent}}\\
    \cline{2-4}
     & \textbf{I1} & \textbf{I2} & \textbf{I3} \\
    \midrule
    IIJ & \makecell{Partially\\disagree} & \makecell{Partially\\agree} & \makecell{Partially\\agree} \\
    \midrule
    CCI & Agree & Agree & Agree \\
    \midrule
    CPM & \makecell{Partially agree} & Agree & Agree \\
    \midrule
    FDC & Agree & Agree & Agree \\
    \midrule
    USI & Agree & Agree & Agree \\
    \midrule
    SDP & Agree & Agree & Agree \\
    \midrule
    DCC & Agree & Agree & Agree \\
    \midrule
    OWI & Agree & Agree & Agree \\
    \midrule
    FCO & Agree & \makecell{Partially\\disagree} & \makecell{Partially\\agree} \\
    \midrule
    ODI & Agree & Agree & Agree \\
    \midrule
    MAI & \makecell{Partially agree} & Agree & Agree \\
    \midrule
    MFI & Agree & Agree & Agree \\
  \bottomrule
\end{tabular}
\end{table}

The adapted TAM questions and their results are shown in Tables~Table~\ref{tab:tam_index} and~\ref{tab:tam}. Due to space restrictions, the index of each question in depicted in Table~\ref{tab:tam_index}. It is possible to observe a strong positive perception, once 25 from 27 questions yield an agreement response. Only T2 and T6 present neutral responses. The positive results on perceived usefulness, ease of use and intention to use indicate that our proposed catalog is helpful and that developers would show willingness to apply it.

\begin{table*}
  \caption{TAM questions index}
  \label{tab:tam_index}
    \begin{tabularx}{\linewidth}{ccX}
    \toprule
    \textbf{Dimension} &
    \textbf{Index} &
    \textbf{Question} \\
    \midrule
    \multirow{4}{*}{Usefulness} & T1 & Being aware of the proposed DI anti-patterns would improve my performance in preventing DI related problems in software systems (i.e. preventing faster) \\
    \cline{2-3}
    & T2 & Being aware of the proposed DI anti-patterns would improve my productivity in preventing DI related problems in software systems (i.e. preventing more and faster) \\
    & T3 & Being aware of the proposed DI anti-patterns would enhance my effectiveness in preventing DI related problems in software systems (i.e. preventing more) \\
    & T4 & I would find the proposed catalog of DI anti-patterns useful in my job \\
    \cline{1-3}
    \multirow{4}{*}{Ease of use} 
    & T5 & Learning to use the proposed catalog of DI anti-patterns would be easy for me \\
    & T6 & I would find it easy to use the proposed catalog of DI anti-patterns to prevent DI related problems in software systems \\
    & T7 & It would be easy for me to become aware of the proposed catalog of DI anti-patterns \\
    & T8 & I would find the proposed catalog of DI anti-patterns easy to apply\\
    \cline{1-3}
    \multirow{1}{*}{Intention to use}
    & T9 & I intend to apply the proposed catalog of DI anti-patterns regularly at work\\
    \bottomrule
  \end{tabularx}
\end{table*}

\begin{table}
\centering
  \caption{Respondents perception over the catalog of DI anti-patterns}
  \label{tab:tam}
    \begin{tabular}{cccc}
    \toprule
    \multirow{1}{*}{\textbf{Index}} &
    \multicolumn{3}{c}{\textbf{Respondent}} \\
    \cline{2-4}
    & \textbf{I1} & \textbf{I2} & \textbf{I3} \\
    \midrule
    T1 & Strongly Agree & Strongly Agree & Strongly Agree \\
    T2  & Agree & Strongly Agree & Neutral \\
    T3 & Strongly Agree & Strongly Agree & Strongly Agree \\
    T4 & Strongly Agree & Strongly Agree & Strongly Agree \\
    T5 & Strongly Agree & Agree & Agree \\
    T6 & Strongly Agree & Strongly Agree & Neutral \\
    T7 & Strongly Agree & Strongly Agree &  Agree \\
    T8 & Strongly Agree & Strongly Agree &  Agree \\
    T9 & Strongly Agree & Strongly Agree & Neutral \\
    \bottomrule
  \end{tabular}
\end{table}

\subsection{Online Survey}
\label{sec:online}

With the lessons learned and the preliminary evidence collected from the application of an interview-based survey described in Section \ref{sec:interview}, we decided to pursue further evidence on the usefulness of the proposed catalog by collecting the opinions of a wider range of developers. Although the interview-based survey provided an in-depth assessment of every aspect of the catalog, the process is not time-effective, taking on average two hours of interview to collect the necessary evidence. Thus, to decrease the time spent on collecting evidence about the usefulness, ease of use, and acceptance of the catalog, but also supporting the respondents with a suitable questionnaire-based survey, a Google forms \footnote{www.google.com/forms/about} survey was used to implement an online questionnaire.

\subsubsection{Design}

We basically converted the survey provided in a form of document to respondents to an online version. Again, the first part of the questionnaire is responsible to gather background information of the respondent. As the respondent has no support from researchers, i.e., it is self-administered, we have decreased the number of information required by summarizing specific questions in a unique one. For instance, instead of inquiring about object-oriented technical skills, we require information about experience in software development. Also, we inquire about expertise in dependency injection in terms of years and level of experience, particularly in the industry. As we aimed to secure that respondents complete the questionnaire, these changes diminish the burden of responding to such a long questionnaire. The background information for the online surveys conducted ahead are shown in Table~\ref{tab:background_index}.

\begin{table}
\centering
  \caption{Background information required for online survey}
  \label{tab:background_index}
    \begin{tabular}{cc}
    \toprule
    \textbf{Information} &
    \textbf{Identifier} \\
    \midrule
     Academic background & B1 \\
    \midrule
    Experience developing software & B2 \\
    \midrule
    Current position & B3 \\
    \midrule
    Dependency injection employment & B4 \\
     \midrule
    Experience with dependency injection & B5 \\
    \bottomrule
  \end{tabular}
\end{table}

For the second part of the questionnaire, we have maintained the same structure. However, during the interviews we observed that, when presenting snippets extracted directly (i.e., with no modifications) from real-world examples, the expert developers asked us about the meaning of classes and attributes presented, deviating the focus of understanding the anti-pattern and completing the survey. Based on this experience, we adapted the code snippets to ease developers' comprehension. These modifications mainly concerned renaming classes, methods, and attributes and improving the code structure for readability. Also, differently from the previous survey described in Section \ref{sec:interview}, we have included the following information for each anti-pattern:

\begin{itemize}
\item A text describing the example of occurrence of the anti-pattern, including the main elements involved, such as attributes and methods
\item A text describing the suggested resolution, including information about the code transformations performed
\end{itemize}

The descriptions about the examples of source code provided were important to allow a further and faster understanding of the code snippets. In the earlier survey (described in Section \ref{sec:interview}), we have provided these descriptions verbally as requested by the interviewees. Furthermore, as we also aim to understand the applicability of a refactoring process in each anti-pattern, a new question was added, which is \textit{"Would you fix the anti-pattern on source code? Why?"}

The new question aims to collect the willingness of the developer to perform an anti-pattern fix on source code. We particularly introduced this question in order to investigate the prospects of future work towards patterns for refactoring DI anti-patterns in source code. Lastly, in the final part of the questionnaire, TAM questions were maintained exactly as found in the first survey.

Again, the final part of the questionnaire concerned the application of the TAM. The complete instrumentation employed in the online survey can be accessed online\footnote{https://zenodo.org/record/3610177}.

\subsubsection{Execution of the Online Expert Survey}


In order to verify the applicability of the new designed survey to be openly available online, we have run a study solely with expert developers. We aimed aimed at comprehending if the form was sufficient to be openly available without guidance and support from researchers conducting the study.

Again, we have relied on the strategy of identifying a sample of the population that would meet our survey design. Extensive experience in industry and software engineering were pre-requisites we addressed. The expert online survey was openly available from July, 10th until October, 29th. In this assessment, a total of 11 respondents were contacted and 6 completed the online questionnaire. We informally contacted the respondents to gather feedback. Most of them described the online survey as "easy to understand and fill", however, filling the survey was described as "time-consuming" due to the extension of the questionnaire. The background of the respondents of the expert online survey is shown in Table~\ref{tab:background_2}. As can be observed, the respondents have strong expertise in software development and applying dependency injection. I3 does not posses a strong experience in DI, however, we consider I3 a valuable respondent due to his extensive experience with software engineering and object-oriented design in industry. 

\begin{table*}
  \caption{Background of respondents of the online survey only with expert practitioners}
  \label{tab:background_2}
  \begin{tabularx}{\linewidth}{XXXX}
    \toprule
    \multirow{2}{*}{\textbf{Information}} &
    \multicolumn{3}{c}{\textbf{Respondent}} \\
    \cline{2-4}
     & \textbf{I1} & \textbf{I2} & \textbf{I3}  \\
    \midrule
    B1 & Master & Master & Master \\
    \midrule
    B2 & 16 years & 20 years & 23 years \\
    \midrule
    B3 & Systems Analyst & Professor & Systems Analyst \\
    \midrule
    B4 & Several projects in industry & Several projects in industry & A project in industry \\
     \midrule
    B5 & 12 years & 15 years & 1 year \\
    \bottomrule
   & \multicolumn{3}{c}{\textbf{Respondent}} \\
    \cline{2-4}
     & \textbf{I4} & \textbf{I5} & \textbf{I6} \\
    \midrule
    B1 & PhD & Master & Master \\
    \midrule
    B2 & 20 years & 19 years & 10 years \\
    \midrule
    B3 & Professor & Project Leader & Team Leader \\
    \midrule
    B4 & Several projects in industry & Several projects in industry & Several projects in industry \\
     \midrule
    B5 & 9 years & 7 years & 5 years \\
    \bottomrule
  \end{tabularx}
\end{table*}

\subsubsection{Execution of the Open Online Survey}

Due to the perception that the survey was large in extension (e.g., some of the invited respondents were taking almost an hour to complete and complained about the large extension), in order to increase the likelihood of having complete answers in the open online survey, we divided the questionnaire into two parts. 

The separation was strongly motivated by a high-level categorization of the anti-patterns. Although this categorization was not made part of the anti-pattern template explained in Section~\ref{sec:catalog}, we believe such separation allowed us to mitigate that a fraction of the respondents were mainly responding about a given specific set of correlated anti-patterns. It is worthy to note such categorization was transparent to the respondents. We explain our reasoning as follows.

By reasoning about the catalog in a meta level, we realized that some anti-patterns presented a similar nature. For instance, some anti-patterns clearly concern violating design principles behind dependency injection (CCI, CPM, FDC, OWI, ODI, and MAI) while others are related to quality attributes such as standardization (FCO, MFI) and performance (IIJ, USI), or architectural issues related to eschewing the use of the DI container (SDP, DCC).

Both parts provide distinct anti-patterns and contain the same amount of anti-patterns per category. Table~\ref{tab:division_online_survey} shows the distribution of anti-patterns in two distinct online surveys. After the pilot study and these adjustments, we were more confident that the online survey could be made available to any respondent.

This time, we pursued a wide dissemination strategy, targeting developers and researchers working with dependency injection with diverse backgrounds and from different companies and communities. We asked key industry representatives to disseminate the questionnaire within their companies and also used social media (LinkedIn, Twitter, and Facebook groups focused on software engineering) to recruit participants.


The online survey was openly available from October, 20th until November, 15th. A total of 9 respondents have provided their opinion on the catalog. The background of the respondents of the open online survey is shown in Table~\ref{tab:background_3}. Unfortunately, reaching developers with substantial or sufficient dependency injection experience and willing to answer a detailed feedback survey on anti-patterns is not trivial. Indeed, based on our multi-phase survey experience, we have limited expectations to achieve significantly larger samples with additional replications. Nevertheless, we considered that the survey executions succeeded in including experienced developers which provided detailed feedback for assessing the usefulness from a practitioner point of view, and opted to invest effort into qualitative analyses.


\begin{table}
\centering
  \caption{DI Anti-patterns distribution over two open online surveys}
  \label{tab:division_online_survey}
  \begin{tabular}{ccc}
    \toprule
    \multirow{2}{*}{\textbf{Category}} 
    & \multicolumn{2}{c}{\textbf{Open Online Survey}}\\
    \cline{2-3}
     & \textbf{1} & \textbf{2} \\
    \midrule
    Performance & IIJ & USI  \\
    \midrule
    Design & \makecell{CCI\\CPM\\FDC} & \makecell{OWI\\ODI\\MAI} \\
    \midrule
    Architecture & SDP & DCC  \\
    \midrule
    Standardization & FCO & MFI \\
  \bottomrule
\end{tabular}
\end{table}

\begin{table}
\centering
  \caption{Background of respondents of the open online survey}
  \label{tab:background_3}
  \begin{tabularx}{\linewidth}{XXXX}
    \toprule
    \multirow{2}{*}{\textbf{Information}} &
    \multicolumn{3}{c}{\textbf{Respondent}} \\
    \cline{2-4}
     & \textbf{I7} & \textbf{I8} & \textbf{I9}  \\
    \midrule
    B1 & Master & Bachelor & Master \\
    \midrule
    B2 & 7 years & 11 years & 10 years \\
    \midrule
    B3 & PhD Student & Developer & Project Manager \\
    \midrule
    B4 & For my own use & Several projects in industry & Several projects in industry \\
     \midrule
    B5 & 2 years & 3 years & 3 years \\
    \bottomrule
   & \multicolumn{3}{c}{\textbf{Respondent}} \\
    \cline{2-4}
     & \textbf{I10} & \textbf{I11} & \textbf{I12} \\
    \midrule
    B1 & Bachelor & Bachelor & Bachelor \\
    \midrule
    B2 & 10 years & 3 years & 3 years \\
    \midrule
    B3 & Developer & Developer & Developer \\
    \midrule
    B4 & Several projects in industry & Several projects in industry & Several projects in industry \\
     \midrule
    B5 & 3 years & 2 years & 3 years \\
    \bottomrule
       & \multicolumn{3}{c}{\textbf{Respondent}} \\
    \cline{2-4}
     & \textbf{I13} & \textbf{I14} & \textbf{I15} \\
    \midrule
    B1 & Bachelor & Bachelor & Bachelor \\
    \midrule
    B2 & 4 years & 5 years & 3 years \\
    \midrule
    B3 & Developer & Developer & Master's Student \\
    \midrule
    B4 & Several projects in industry & Several projects in industry & Several projects in industry \\
     \midrule
    B5 & 4 years & 1 year & 1 year \\
    \bottomrule
  \end{tabularx}
\end{table}

\subsubsection{Results}
\label{subsec:online_expert_survey}

The results of the first six respondents, which were previously selected to participate in the online survey, on the proposed DI anti-patterns, are exhibited in Table~\ref{tab:perception_preliminary_online_survey}. The results of the openly available online survey are exhibited in Tables~\ref{tab:perception_openly_online_survey_d1} and~\ref{tab:perception_openly_online_survey_d2}. 

\begin{table*}
  \caption{Perception over the DI anti-patterns in the expert online survey}
  \label{tab:perception_preliminary_online_survey}
  \begin{tabular}{ccccccc}
    \toprule
    \multirow{2}{*}{\makecell{\textbf{DI} \\ \textbf{Anti-Pattern}}}
    & \multicolumn{6}{c}{\textbf{Respondent}}\\
    \cline{2-7}
     & \textbf{I1} & \textbf{I2} & \textbf{I3} & \textbf{I4} & \textbf{I5} & \textbf{I6} \\
    \midrule
    IIJ & Agree & \makecell{Partially\\disagree} & \makecell{Partially\\agree} & Agree & \makecell{Partially\\agree} & \makecell{Partially\\disagree} \\
    \midrule
    CCI & Agree & Agree & \makecell{Partially\\agree} & Agree & Agree & \makecell{Partially\\agree} \\
    \midrule
    CPM & Agree & Neutral & Agree & Agree & Agree & Agree \\
    \midrule
    FDC & \makecell{Partially\\agree} & Neutral & Agree & Agree & Agree & Agree \\
    \midrule
    USI & Agree & Agree & Agree & Agree & Agree & Agree \\
    \midrule
    SDP & \makecell{Partially\\disagree} & Agree & Agree & Agree & Agree & Agree \\
    \midrule
    DCC & \makecell{Partially\\disagree} & \makecell{Partially\\agree} & \makecell{Partially\\agree} & Agree & Agree & Agree \\
    \midrule
    OWI & \makecell{Partially\\agree} & \makecell{Partially\\agree} & Agree & Agree & Agree & Agree \\
    \midrule
    FCO & \makecell{Partially\\agree} & \makecell{Partially\\agree} & Disagree & Agree & Agree & Agree \\
    \midrule
    ODI & \makecell{Partially\\agree} & Agree & Agree & Agree & Agree & Agree \\
    \midrule
    MAI & Agree & Agree & Agree & Agree & Agree & Agree \\
    \midrule
    MFI & Agree & Agree & Agree & Agree & Agree & Agree \\
  \bottomrule
\end{tabular}
\end{table*}

\begin{table}
\centering
  \caption{Perception over the DI anti-patterns in openly online survey 1}
  \label{tab:perception_openly_online_survey_d1}
  \begin{tabular}{ccccc}
    \toprule
    \multirow{2}{*}{\makecell{\textbf{DI} \\ \textbf{Anti-Pattern}}}
    & \multicolumn{4}{c}{\textbf{Respondent}}\\
    \cline{2-5}
     & \textbf{I7} & \textbf{I8} & \textbf{I9} & \textbf{I10} \\
    \midrule
    IIJ & Agree & \makecell{Partially\\disagree} & \makecell{Partially\\agree} & Agree  \\
    \midrule
    CCI & Agree & Agree & \makecell{Partially\\disagree} & Agree  \\
    \midrule
    CPM & Neutral & \makecell{Partially\\agree} & Agree & \makecell{Partially\\agree} \\
    \midrule
    FDC & \makecell{Partially\\disagree} & Agree & Agree & \makecell{Partially\\agree} \\
    \midrule
    SDP & \makecell{Agree} & Agree & \makecell{Partially\\agree} & Neutral  \\
    \midrule
    FCO & Agree & Agree & Agree & Neutral \\
  \bottomrule
\end{tabular}
\end{table}

\begin{table}
\centering
  \caption{Perception over the DI anti-patterns in openly online survey 2}
  \label{tab:perception_openly_online_survey_d2}
  \begin{tabular}{cccccc}
    \toprule
    \multirow{2}{*}{\makecell{\textbf{DI} \\ \textbf{Anti-Pattern}}}
    & \multicolumn{5}{c}{\textbf{Respondent}}\\
    \cline{2-6}
     & \textbf{I11} & \textbf{I12} & \textbf{I13} & \textbf{I14} & \textbf{I15} \\
    \midrule
    USI & \makecell{Partially\\disagree} & Agree & Agree & Agree & Agree  \\
    \midrule
    DCC & Agree & \makecell{Partially\\agree} & Disagree & Neutral & \makecell{Partially\\agree}  \\
    \midrule
    OWI & Agree & Agree & Disagree & Agree & \makecell{Partially\\agree}  \\
    \midrule
    ODI & \makecell{Partially\\agree} & Agree & Agree & Agree & Agree \\
    \midrule
    MAI & Agree & Neutral & Agree & Agree & Agree \\
    \midrule
    MFI & Agree & Disagree & Agree & Agree & Agree \\
  \bottomrule
\end{tabular}
\end{table}

It is important to highlight that only 4 anti-patterns (IIJ, SDP, DCC, and FCO) present disagreements. However, SDP and DCC present disagreements from a respondent with the highest rate of disagreements (I1). By conducting a qualitative analysis on his responses, we have observed that he considers the \textit{Service Locator} pattern a good practice when it comes to integrate several frameworks in a software system, although he considers the strong dependence on these types of classes a drawback (SDP). We argue that there are better approaches to refrain the system to be coupled to a singleton object, such as the use of \textit{Provider} classes and \textit{Producer} methods, which were explained along the instrumentation. Also, he asserts that the use of direct container calls does not yield drawbacks to the application, since he "never experienced the change of the framework of a project in production." We argue that the proposal on addressing direct container call as an anti-pattern is related to the increased effort in future maintenance activities on the application. Besides, the disagreements on IIJ and FCO are results that we already expected, since the interview-based survey provided the same insights.

On the other side, we observed a positive perception over the DI anti-pattern instances. From 72 enquiries, only 4 disagreements and only 2 neutral responses are observed. It is noteworthy to mention that some neutral and disagreements responses do not come from the anti-pattern instance conjectured, but rather the resolution provided. The qualitative analysis as follows will go over these instances.

For the first openly online survey, which covers anti-patterns IIJ, CCI, CPM, FDC, SDP, and FCO, only 3 (partial) disagreements are observed. In general, from 24 enquiries, 18 responses provided agreements over the instances. Again, we observed that some partial disagreements aimed at the resolution provided and not the instance of anti-pattern presented. 

Due to the lack of expertise of the respondents, as expected, the comments regarding the anti-patterns were not as substantial as the ones provided by the expert developers in the previous survey. Besides, we assert that it is worthy to consider their points to strengthen our evidence on the validity of the proposed DI anti-pattern instances.

Regarding the answers on the willingness of the respondents on fixing the candidate DI anti-pattern, because of the format of open-ended question, a qualitative analysis was conducted over each response. The overall perceptions are presented in Table~\ref{tab:willingness_on_fixing} and described hereafter. Some respondents (\textbf{I1} and \textbf{I5}) have provided their opinion in Portuguese, so in order to explain the results, we provide a direct translation to English.

\begin{table*}
\centering
  \caption{Willingness to fix DI Anti-Patterns}
  \label{tab:willingness_on_fixing}
  \begin{tabular}{cccccc}
    \toprule
    
    \multirow{2}{*}{\makecell{\textbf{DI} \\ \textbf{Anti-Pattern}}}
    
    & \multicolumn{4}{c}{\textbf{Williness to Fix DI Anti-Pattern}} \\
    
    \cline{2-5}
     & \textbf{Yes} & \textbf{Yes, conditionally} & \textbf{No} & \textbf{Do not know} \\
    
    \midrule
    
    IIJ  & 30.0\% (3/10)   & 30.0\% (3/10) & 40.0\% (4/10) & -  \\ \midrule
    CCI  & 60.0\% (6/10)   & 20.0\% (2/10) & 20.0\% (2/10) & -  \\ \midrule
    CPM  & 70.0\% (7/10)   & 10.0\% (1/10) & 10.0\% (1/10) & 10.0\% (1/10)  \\ \midrule
    FDC  & 90.0\% (9/10)   & -             & 10.0\% (1/10) & -  \\ \midrule
    USI  & 100.0\% (10/10) & -             & -             & -  \\ \midrule
    SDP  & 70.0\% (7/10)   & -             & 30.0\% (3/10) & -  \\ \midrule
    DCC  & 54.5\% (6/11)   & 9.1 (1/11)    & 18.2\% (2/11) & 18.2 (2/11)\%  \\ \midrule
    OWI  & 90.9\% (10/11)  & -             & 9.1 (1/11)\%  & -  \\ \midrule
    FCO  & 70.0\% (7/10)   & -             & 30.0\% (3/10) & -  \\ \midrule
    ODI & 81.2\% (9/11)   & -             & 9.1 (1/11)\%  & 9.1 (1/11)\%  \\ \midrule
    MAI & 90.9\% (10/11)  & -             & 9.1 (1/11)\%  & -  \\ \midrule
    MFI & 100.0\% (11/11) & -             & -             & -  \\ 
    
  \bottomrule
\end{tabular}
\end{table*}

\subsubsection{Qualitative Analysis}

In the following we provide a thorough discussion over the opinions expressed by the respondents for each proposed anti-pattern.

\noindent \textbf{IIJ.} Most of the respondents answered that they would fix IIJ unconditionally (30\%) or in certain conditions (30\%). I4 said why she would fix IIJ \textquote{the injection of unnecessary dependencies demands computing time}. The condition stated to fix IIJ was the overload of computational resources in the system. The respondent I3 said \textquote{I would only fix it if the code in the example is too heavy to be loaded eagerly}. Finally, 40\% of respondents would not intend to fix IIJ because of the risk of introducing bugs could not pay-off the performance or computational resource gains. The respondent I9 said \textquote{No (would not fix IIJ), the overload is small and should not be a problem}, while I2 explained the kind of bug that can be introduced by fixing IIJ \textquote{... the application can start with unresolved injections which may be potentially broken, leveraging the problem to run-time instead of startup time}.

\textit{Takeaway.} This opens up opportunities to better devise techniques, tools, and frameworks that allow developers to effectively reason about the dependencies at start up, informing users of misconfigurations or configurations that are error-prone in runtime. In sum, they want to avoid misuses of resources but they feel fear of introducing problems, which suggests current tools are not comprehensive in this aspect.

\textbf{CCI.} Most of respondents intend to fix CCI (60\%) unconditionally or would fix it in certain conditions (20\%). The reasons stated for fixing CCI were that CCI affects inversion of control and its solution allows transversal requirements and avoid coupling and refactoring. I2 respondent said \blockquote{Yes, the injection being based on interfaces allows for dynamic proxying, and automation of transactional control
and similar transversal requirements}. The conditions to fix CCI was the real necessity of decoupling classes and an opportunity for refactoring. I3 respondent said \textquote{I would only fix it if there is a real need for interface decoupling. I see no point in having a lot of interfaces and only one concrete class for each one of them}. Only 20\% of respondents would not intend to fix CCI because it could be a low priority change or a not needed change for small systems. I9 respondent said \textquote{No, there is no need to make an interface for small systems. If it is needed you can generate the interface afterwards upon need}.

\textit{Takeaway.} The introduction of interface-oriented design towards achieving DIP is positively seen by practitioners when it comes to larger systems. This suggests they do not see the benefit entailed by DIP in small systems, even though DI is employed.

\textbf{CPM.} The majority of respondents informed that they would fix CPM unconditionally (70\%) or with conditions (10\%). The reasons stated to fix CPM were improving cohesion and comprehensibility of code as well as reducing complexity of it. Respondent I6 said \blockquote{Yes. It seems the code below @Produce is doing more than expected so the programmers need to do more investigation than needed to understand it}. The condition stated to fix CPM was if it generate performance problems in the system. Only one respondents (10\%) would not fix CPM. The reasons for not fixing CPM were the complexity and cost of change. I1 said \textquote{I think the correction of this comes from a deeper refactoring that is not simple to do and expensive when talking about a legacy system}. One respondent (10\%) did not defined if would fix or not CPM because she did not identify CPM as a DI problem, but as a general design problem. I2 respondent said \textquote{The problem depicted on CPM seems to me to be caused by bad architectural design, and not a bad practice of dependency injection itself}.

\textit{Takeaway.} Although the respondents shown willingness to refactor such instances, some of them pointed out that this anti-pattern is more related to general design and architectural problems in the system, e.g., bad design choices earlier in the project that would lead to this kind of problem. The lack of a statistical significance makes this anti-pattern less appealing in practice, which does not diminish the importance of promoting such avoidance.

\textbf{FDC.} Almost all respondents (90\%) informed that they unconditionally intend to fix FDC. The reasons stated for fixing FDC were to improve modularity, cohesion and maintainability of code. Respondent I6 said \textquote{it seems class D is doing to much and it will be hard
to maintain} and I3 said \textquote{The code should be simpler and modularized}. However, one respondent (10\%) would not fix FDC because she identify it as an architectural problem listed in other anti-pattern catalog. Respondent I2 said \textquote{this doesn't seem to me as a dependency injection problem, but really an architectural design flaw, already addressed by basic object orientation practices and most common pattern catalogs}.

\textit{Takeaway.} Respondents mainly agree that source code with significant introduction of injections may be harmful to the system in the long term, diminishing the ability of the system to evolve.

\textbf{USI.} All respondents (100\%) stated that would fix USI without conditions. The reasons for that were for cleaning the code and improving performance. Respondent I4 said \textquote{because the resulting code if easier do understand and more performatic}. Furthermore, one respondent suggested that USI could be alerted by default in the developer IDE. Respondent I3 said \textquote{A good well configured IDE can do this or warn me to do it}.

\textit{Takeaway.} IDEs do not alert users about non used injection. They usually understand an injected dependence as something used by the system, even though might not be.

\textbf{SDP.} SDP is the use of a static dependence provider. Most of respondents (70\%) informed that they would fix SDP. The reasons stated for fixing SDP were a better decoupling and flexibility of code. Respondent I4 said \textquote{the resulting code becomes more flexible and generic}, while I2 said \textquote{The fix on SDP shows the most decoupled organization of code}. Besides of that, some respondents (30\%) would not fix SDP because SDP is not considered a big problem or could not have this problem without using a framework. The respondent I5 said \textquote{Since I don't currently use a DI framework / lib, would have difficulty seeing this lack of standardization in solution to obtain dependency in two different ways}. Furthermore, one respondent (I10) suggested a different solution for SDP  \textquote{You don’t need to use the ServiceLocator to get the IDataSource each time you have to use it. You can create a global object and you can control it to call just once. In this case, DI doesn’t make such a huge difference to me}.

\textit{Takeaway.} The responses suggest that some developers are not aware of the anomalies DI is expected to avoid, suggesting the introduction of anomalies by not following DI principles. This is observed in responses from less experienced developers.

\textbf{DCC.} Most of respondents would fix DCC (54.5\%) unconditionally, and one conditionally (9.1\%). The main reason stated for fixing DCC was to reduce coupling with specific frameworks. Respondent I6 said \textquote{the main goal of using DI is not depending directly on the injected objects. Also, depending on a framework would create the same kinds of problems}. The condition stated for fixing DCC was in case of DI framework changing. Respondent I1 said \textquote{I've never seen the framework change in practice for a project in production}. The other respondents (18.2\%) would not fix DCC because it could introduce problems and complexity. Respondent I13 said \textquote{This may be used to improve performance, only loading some dependency when it is absolutely needed. The extra provider will add unnecessary complexity, only hiding the original intention}. Other two (18.2\%) respondents do not know exactly if they would fix it or not. They have doubts if a provider or another method could be used to encapsulate the dependency. Respondent I12 said \textquote{Maybe. In this case I would have used a service locator, or at least encapsulate the dependency in a method of the provider object}.

\textit{Takeaway.} It is clear from the less experienced respondents that they do not understand direct coupling to internal framework implementation as harmful, which indicates some illiteracy on dependency injection principles and objectives in a system. In opposition, expert developers primarily agree that such coupling is harmful in the long term and should only be employed in special circumstances, such as tying dependencies of different frameworks together. 

\textbf{OWI.} Almost all respondents (90.9\%) would fix OWI unconditionally. The reasons stated for fixing OWI were to eliminate unnecessary code and reduce coupling. Respondent I14 said \textquote{yes, as it eliminates unnecessary code}, while I5 said \textquote{Yes, refactoring decreases coupling}. However, one respondent (9.1\%) would not fix it because the violation of encapsulation would be needed when integrating with code that does not have access to the container. Respondent I13 said \textquote{No. This situation is necessary when integrating third-party code that will not have access to the DI container. It also allows a function or method to receive different implementations of a parameter}.

\textit{Takeaway.} This suggests OWI is only necessary in specific cases where framework integration is needed. This suggests the tools should provide more disciplined ways to achieve it without hurting DI principles.

\textbf{FCO.} Most of the respondents (70\%) would unconditionally fix FCO. The main reasons stated for fixing FCO were to reduce coupling with frameworks. Respondent I6 said \textquote{Yes, for me it makes no sense to depend on the DI framework. We use DI to reduce coupling so depending on a framework all around the code does not make sense}. However, three of them (30\%) would not fix FCO because either they think that for small projects it is not a problem or that there are benefits to coupling to certain frameworks as Spring. Respondent I3 said \textquote{No, it is not a big problem for small projects where it is highly unlikely to change the framework}. Respondent I10 said \textquote{In this specific case, we are talking about a huge framework that brings lots of benefits to the application (Spring). So, for me it is worth even if it implies in higher coupling to the framework specifics}.

\textit{Takeaway.} The responses indicate that framework coupling is less appealing on medium-to-large projects, due to the intrinsic complexity on maintenance entailed in the long run.

\textbf{ODI.} Almost all respondents (81.8\%) would fix ODI unconditionally. The reasons stated for fixing ODI were to reduce complexity, maintain encapsulation and to prevent problems to propagate to other parts of the architecture. Respondent I2 stated \textquote{I would fix ODI to protect the injection and the whole architectural integrity}. However, one respondent (9.1\%) had doubts on whether he would fix ODI and another one (9.1\%) does not intend to fix it. The reason stated for that concerned doubts about the utility of fixing ODI. Respondent I1 said \textquote{Usually I would not do it, but I can see cases where it can be useful to provide the possibility of changing the default implementation injected to another one.}. Respondent I11 said \textquote{No. This kind of occurrence may be a problem or not, this can vary according to the architecture}.

\textit{Takeaway.} It is clear that ODI harms the encapsulation principle of object-oriented programs, as extensively agreed by respondents. However, it is worthy to point out that this may vary according to the architecture, as explicitly stated by one of the respondents. Although this respondent did not elaborate on the thought regarding this matter, we presume that some applications may require some level of encapsulation breaking in order to tie dependencies across system's modules.

\textbf{MAI.} MAI concerns assigning an injected instance to more than one attribute. Almost all respondents (90.9\%) would fix MAI unconditionally. The reasons stated for that were to simplify the code and avoid bugs. Respondent I4 said \textquote{yes, as it simplifies the code without loosing flexibility}, while I5 said \textquote{Yes, because this situation can lead to confusing defects for debugging}. However, one respondent (I12) (9.1\%) would not fix MAI because he did not agree with the solution given by the catalog.
He mentioned that \textquote{I don’t understood why the solution is right, both of them look wrong}, without providing further explanations.

\textit{Takeaway.} The perception that multiple assigned injection is harmful to the application code is prevalent. IDEs usually do not alert users about such practice. Some less experienced developers again expressed some wrong idealized assumptions about assigning dependencies by a DI container.

\textbf{MFI.} All respondents (100\%) would fix MFI. The reasons stated for fixing MFI were to better understand and maintain the code of the application. Respondent I4 said \textquote{Yes, as it makes the code easier to understand and maintain}, while I2 said \textquote{I would fix MFI to maintain the overall convention and organization of the project}.

\textit{Takeaway.} The positive perception regarding fixing the anti-pattern is absolute. Again, IDEs do not alert users about such practice.

\subsection{Discussion}

Although several responses showed a partial agreement, by verifying their responses on the willingness of fixing the anti-pattern, the respondents expressed an agreement with the anti-pattern, only expressing concerns over a conjectured scenario. Thus, the qualitative analysis was very important to uncover this kind of pattern in the responses. In addition, adding the question over fixing the anti-patterns also allowed the collection of responses of this nature. The results strengthen our confidence that the proposed catalog is important for development activities involving the application of DI.

After responding on the DI anti-patterns, the respondents were enquired about the utility of the catalog based on TAM facets. The adapted TAM questionnaire and the results are exhibited in the Table~\ref{tab:tam_1}. These were ordered and positioned according to the results previously depicted. The Index column of both tables address the same questions depicted in Table \ref{tab:tam_index}. 

In the online expert survey, a strong positive perception is observed. 31 of 36 yield an agreement response, on which 3 of the 5 neutral responses came from a specific respondent. 

\begin{table}
\centering
  \caption{Respondents perception over the catalog of DI anti-patterns (I1-I6)}
  \label{tab:tam_1}
    \begin{tabular}{cccc}
    \toprule
    \multirow{1}{*}{\textbf{Index}} &
    \multicolumn{3}{c}{\textbf{Respondent}} \\
    \cline{2-4}
    & \textbf{I1} & \textbf{I2} & \textbf{I3} \\
    \midrule
    T1 & Strongly Agree & Agree & Agree \\
    T2 & Strongly Agree & Agree & Agree \\
    T3 & Strongly Agree & Agree & Agree \\
    T4 & Strongly Agree & Neutral & Agree \\
    
    T5 & Strongly Agree & Agree & Agree \\
    T6 & Strongly Agree & Neutral & Agree \\
    
    T7 & Strongly Agree & Agree & Agree \\
    T8 & Strongly Agree & Agree & Agree \\
    T9 & Neutral & Neutral & Neutral \\
    \bottomrule
    & \textbf{I4} & \textbf{I5} & \textbf{I6} \\
    \midrule
    T1 & Strongly Agree & Agree & Agree \\
    T2 & Strongly Agree & Agree & Strongly Agree \\
    T3 & Strongly Agree & Agree & Strongly Agree \\
    T4 & Strongly Agree & Agree & Strongly Agree \\
    
    T5 & Strongly Agree & Agree & Strongly Agree \\
    T6 & Strongly Agree & Agree & Agree \\
    
    T7 & Strongly Agree & Agree & Neutral \\
    T8 & Strongly Agree & Agree & Agree \\
    T9 & Strongly Agree & Agree & Agree \\
    \bottomrule
  \end{tabular}
\end{table}




Regarding the openly online survey, the adapted TAM questionnaires and their results are exhibited in the Tables~\ref{tab:tam_2} and~\ref{tab:tam_3}. Again, these were ordered and positioned according to previous results and the Index column of both tables address the same questions depicted in Table \ref{tab:tam_index}. 
From 81 TAM questions, only 17 did not yield an agreement response. We consider the results positive and strengthen our confidence on the usefulness, ease of use and intention to use of the proposed catalog. Most importantly, the results suggests that the catalog is helpful and that developers show willingness to apply it.

\begin{table*}
\centering
  \caption{Respondents perception over the catalog of DI anti-patterns (I7-I10)}
  \label{tab:tam_2}
    \begin{tabular}{ccccc}
    \toprule
    \multirow{1}{*}{\textbf{Index}} &
    \multicolumn{4}{c}{\textbf{Respondent}} \\
    \cline{2-5}
    & \textbf{I7} & \textbf{I8} & \textbf{I9} & \textbf{I10} \\
    \midrule
    T1 & Strongly Agree & Strongly Agree & Agree & Strongly Agree \\
    T2 & Strongly Agree & Strongly Agree & Agree & Strongly Agree \\
    T3 & Strongly Agree & Strongly Agree & Agree & Strongly Agree \\
    T4 & Agree & Strongly Agree & Strongly Agree & Strongly Agree \\
    
    T5 & Disagree & Neutral & Strongly Agree & Strongly Agree \\
    T6 & Disagree & Neutral & Strongly Agree & Strongly Agree \\
    
    T7 & Strongly Agree & Neutral &  Agree & Strongly Agree \\
    T8 &  Disagree & Agree & Strongly Agree & Strongly Agree \\
    T9 & Neutral & Agree & Agree & Strongly Agree \\
    \bottomrule
  \end{tabular}
\end{table*}

\begin{table*}
\centering
  \caption{Respondents perception over the catalog of DI anti-patterns (I11-I15)}
  \label{tab:tam_3}
    \begin{tabular}{cccccc}
    \toprule
    \multirow{1}{*}{\textbf{Index}} &
    \multicolumn{5}{c}{\textbf{Respondent}} \\
    \cline{2-6}
    & \textbf{I11} & \textbf{I12} & \textbf{I13} & \textbf{I14} & \textbf{I15} \\
    \midrule
    T1 & Agree & Agree & Strongly Agree & Agree & Strongly Agree \\
    T2 & Agree & Agree & Strongly Agree & Agree & Strongly Agree \\
    T3 & Agree & Agree & Strongly Agree & Agree & Agree \\
    
    T4 & Strongly Agree & Neutral & Neutral & Neutral & Strongly Agree \\
    
    T5 & Agree & Neutral & Neutral & Agree & Agree \\
    T6 & Agree & Agree & Agree & Agree & Agree \\
    
    T7 & Agree & Agree &  Neutral & Agree & Agree \\
    T8 & Strongly Agree & Disagree & Neutral & Agree & Agree \\
    T9 & Strongly Agree & Neutral & Disagree & Neutral & Agree \\
    \bottomrule
  \end{tabular}
\end{table*}
\section{Updating the Catalog}
\label{sec:embracing}

Based on the opinion of the respondents collected throughout the application of the three surveys (interview-based expert survey on Section \ref{sec:interview}, online expert survey, and openly available online survey on Section \ref{sec:online}), along with comments gathered from the evaluation members, we realized the catalog needed a new version.

The new version of the catalog of DI anti-patterns should embrace the reflections made from the point of view of developers over the catalog. The main changes are summarized in Table \ref{tab:summary_changes_catalog} and the updated version of the catalog can be found online\footnote{https://zenodo.org/record/4679322}. 
In sum, the input from developers corroborate that the catalog is useful in practice, but to allow for appropriate reasoning of thought about the applicability of the DI anti-patterns, further details about the anti-pattern and associated refactoring problem context should be provided.

\begin{table*}
 \caption{Summary of updates in the catalog}
 \begin{tabularx}{\textwidth}{c>{\hsize=0.7\hsize}YY}
  \toprule
 \textbf{Source} & \textbf{Problem/Proposition} & \textbf{Argument/Solution} \\
 \midrule
 IIJ & Whether IIJ is dependent on the problem being solved & Reinforce the problem context where this anti-pattern is applied. \\
 \midrule
 CCI & In cases of small systems, an interface can be created afterwards upon need & Small systems might not benefit from an interface-oriented design in some cases. The updated catalog reinforces this aspect. \\
 \midrule
 CPM & Whether CPM is just an instance of long method bad smell & Long method bad smell is applied to any generic method. CPM, however, is applied only when the provider method performs activities outside of its core scope, which is providing a dependence instance. The updated catalog reinforces this aspect. \\
 \midrule
 FDC & Whether FDC is just an instance of god class bad smell & FDC concerns the injection of a substantial number of dependencies in a class. This anti-pattern is primarily concerned over injected instances that are often introduced by developers without reasoning over the increased dependence of the class of other components. The updated catalog reinforces this aspect. \\
 \midrule 
 USI & Whether USI is an anomaly specifically related to DI & The problem is more related to current integrated development environments (IDEs) which, once annotated, even  though the attribute is not used, do not warn the developer about the issue. The updated catalog reinforces this aspect. \\
 \midrule
 SDP & In cases where different frameworks must be integrated in a single project, a "true" singleton can be used to provide dependencies, even though incurring in strong coupling observed in classes of the project & The anti-pattern does not assert about these types of situations. The updated catalog reinforces this aspect.  \\
 \midrule
 DCC & There are cases in which dependencies may be dynamically resolved at runtime with the support of the DI container & In these cases, the catalog advocates to wrap the container call in a provider class \\
 \midrule
 OWI & In cases where an injected object needs to be passed to a component (or third-party solution) that lives outside the container, the anti-pattern is a solution & Reinforce the problem context. \\
 \midrule
 FCO & The likelihood of changing a previously defined framework in a project is low, customers may not be willing to cover the costs of such change & Reinforce that the decision of relying on the specification is better applied in the context of new software projects \\
 \midrule 
 Structural definition & The lack of explicitly arguing about the context of an anti-pattern made respondents skeptical about its applicability & Introduce the context element in the definition structure \\
 \bottomrule
\end{tabularx}
\label{tab:summary_changes_catalog}
\end{table*}

\section{Threats to Validity}
\label{sec:threats}

The threats to validity are organized according to Wohlin et al. \cite{wohlin:12}.

\subsection{Internal Validity} 

\textbf{Survey.} Regarding the first survey, the interview-based approach was used specifically to clarify any doubts regarding the questions and the proposed catalog, not biasing respondents towards their agreement. It served as a preliminary assessment of the structure of the catalog and the instrumentation employed. Regarding the online survey, we provided more information regarding the source code examples to assist the respondents towards fulfilling the online questionnaire.

\textbf{Tool.} The tool built to flag instances of anti-patterns may miss some instances in the source code, since every software project may show different implementation characteristics. To mitigate this threat, we evaluated the DIAnalyzer tool regarding relative recall and precision. We believe that we have identified most of the DI anti-patterns in source code of the analyzed software projects. In addition, we double-checked the findings with the support of an independent researcher, which carried out an additional manual validation of the instances detected by the tool.

\subsection{External Validity}

\textbf{Survey.} Regarding the interview-based survey, as we planned to conduct a limited amount of interviews with a limited amount of subjects, the instrumentation is available for external replications. Regarding subject representatives, we selected experienced developers from three different organizational units. Regarding the instrumentation, we peer-reviewed the material before presenting it to the subjects. Regarding the expert online survey, although representatives that answered our survey are from the same institution, we assert that this does not yield a threat since they are lecturers and do not work closely. For the online instrumentation, we also used a peer-review process to employ the division of the survey into two distinct surveys. 

\textbf{Catalog of Anti-patterns.} Although we targeted Java software projects for characterizing and detecting DI anti-patterns, we believe most of the proposed DI anti-patterns may apply to projects implemented in other programming languages, especially those following the object-oriented paradigm. For instance, declarative constructs through annotations are also found in other object-oriented programming languages like C\# and Python. Nevertheless, our empirical investigations were scoped to Java software projects and we conservatively relate our findings to this contexts. Further investigations are required to extend the validity of the findings to other contexts.


\textbf{Open-source Analysis.} Albeit selecting projects with different number of LOC and commits, most of them are implemented using the Spring framework. However, the risk of leaning the findings towards a specific framework is mitigated because one of the closed-source projects analyzed employs Guice as DI framework. In addition, our findings were verified in open-source and closed-source systems from industrial settings, which strengthen the practical relevance of the catalog of DI anti-patterns.

\subsection{Construct Validity and Reliability} 

\textbf{Survey.} Our qualitative analysis relies simple Likert scale agreements on the anti-patterns and the TAM statements. We reinforce the interpretation of the results providing argumentation based on open text answers. TAM is a widely employed tool to measure the perceived acceptance of technology propositions.

\textbf{Open-source Analysis.} Since there is no benchmarking dataset for DI anti-patterns, we built an oracle data-set by ourselves in order to evaluate DIAnalyzer. The data-set was built and verified by two independent researchers. The process for building the oracle is similar to Chen and Jiang~\cite{chen:17} work. Lastly, the oracle does not contain instances of CPM. We believe this does not undermine our findings, since this specific anti-pattern did not appeared in any analyzed project.
\section{Concluding Remarks}
\label{sec:conclusion}

The goal of this research concerns addressing the lack of guidance on how to effectively detect, analyze, and remove DI anti-patterns from source code elements.

First, we investigated the literature on DI with the aim of identifying existing documentation on DI anti-patterns. We observed that academic literature does not properly cover DI anti-patterns in source code. Industry-oriented publications, on the other hand, are too generic and fail to provide empirical evidence on the practical relevance of their propositions.

Hence, we applied two methodological approaches to derive an initial catalog of Java DI anti-patterns. Based on observations of bad implementation practices related to the employment of DI in closed-source projects in past work experiences (inductive approach) and conjecturing over a set of instances that harm the principles behind DI (deductive approach), namely, DIP and IoC, and object-oriented design principles, such as GRASP and SOLID, an initial effort towards documenting a catalog of Java DI anti-patterns was taken.

Second, motivated by our proposition over a candidate set of DI anti-patterns, we designed a static analysis tool called DIAnalyzer to automatically flag and report instances of anti-patterns from source code. An evaluation carried out on DIAnalyzer revealed that the tool is reliable and can effectively retrieve instances of DI anti-patterns from the source code. Then, we applied the tool to a set of software systems, both open and closed-source. The investigation revealed that the DI anti-patterns are general and occur within different projects.

Lastly, we designed a study to analyze the acceptance and usefulness of the catalog from the point of view of expert developers. In order to allow an initial in-depth evaluation of our proposed catalog, an interview-based survey was undertaken to mitigate risks related to the design of the instrumentation and description of each anti-pattern. Feedback gathered from the expert developers regarding the instrumentation was used as input for the next step. Then, in order to scale our evaluation, we designed an online version of our survey. The results indicate that the catalog is perceived by practitioners as relevant and useful. Moreover, based on the collected feedback, we built an updated version of the DI anti-patterns catalog. 

Based on our investigations we are confident that the resulting catalog can be useful to help practitioners in avoiding DI anti-patterns, improving the resulting source code quality. Furthermore, the catalog and its evaluations provide insights into new IDE features that could be provided to developers to avoid violating DI related principles.



\section*{Acknowledgement}
This work was financially supported by CNPq. We are also grateful to the practitioners that collaborated to this work, gently providing their insights.

\bibliographystyle{cas-model2-names}

\bibliography{cas-refs}


\end{document}